\documentclass[twocolumn]{aastex631}

\usepackage{booktabs, hyperref, comment, tabularx, ragged2e}
\usepackage[T1]{fontenc}
\usepackage{tabularx}

\begin{document}

\title[M31N 2008-12a]{Far Ultraviolet Variability on Timescales of a Few hours in the Recurrent Nova M31N~2008-12a: Origin and Implications}

\author[0000-0001-7570-545X]{Judhajeet Basu}
\affiliation{Physics Department, Indian Institute of Technology Bombay, Powai, Mumbai, 400076, India}
\affiliation{Indian Institute of Astrophysics, II Block, Koramangala, Bengaluru 560034, India}

\author[0000-0002-3927-5402]{Sudhanshu Barway}
\affiliation{Indian Institute of Astrophysics, II Block, Koramangala, Bengaluru 560034, India}

\author[0000-0003-3533-7183]{G.C. Anupama}
\affiliation{Indian Institute of Astrophysics, II Block, Koramangala, Bengaluru 560034, India}

\author[0009-0000-5909-293X]{Shatakshi Chamoli}
\affiliation{Indian Institute of Astrophysics, II Block, Koramangala, Bengaluru 560034, India}
\affiliation{Pondicherry University, R.V. Nagar, Kalapet, Pondicherry 605014, India}

\author[0000-0001-6952-3887]{Kulinder Pal Singh}
\affiliation{Department of Physical Sciences, Indian Institute of Science Education and Research Mohali, 140306, Punjab, India}

\author[0000-0002-6112-7609]{Varun Bhalerao}
\affiliation{Physics Department, Indian Institute of Technology Bombay, Powai, Mumbai, 400076, India}

\author[0000-0002-0000-1543]{M. Pavana}
\affiliation{Indian Institute of Astrophysics, II Block, Koramangala, Bengaluru 560034, India}

\author[0000-0002-7942-8477]{Vishwajeet Swain}
\affiliation{Physics Department, Indian Institute of Technology Bombay, Powai, Mumbai, 400076, India}

\begin{abstract}

The fate of the accretion disk during nova eruptions remains an open question, particularly in rapidly recurring novae, where high accretion rates and low ejecta masses may allow the disk to survive the eruption. M31N~2008-12a, the most rapidly recurring nova currently known, provides an ideal laboratory for investigating this process. We present high-cadence \textit{AstroSat}/UVIT observations obtained during eruptions between 2019 and 2023. Complementary multiwavelength observations of the 2023–2025 eruptions are also presented, which are broadly consistent with prior outbursts. The main result is the early onset of intrinsic far-ultraviolet variability on a timescale of a few hours, $\sim$4--5 days after eruption, which is of astrophysical origin. A periodicity search suggests a tentative signal at $4.60 ^{+0.25}_{-0.15}$\,hr at 2.3$\sigma$ confidence, detected in high-cadence UVIT/FUV observations. We show that a Keplerian orbit of this timescale lies not only within the Roche lobe of the white dwarf in M31N~2008-12a, but also within the circularisation radius. A disk-related origin is, thus, physically plausible. The presence of this UV variability so soon after eruption suggests that accretion structures may already exist during the early post-eruption phase, supporting the disk-survival hypotheses. If confirmed by future observations, this method may provide a new dynamical probe of accretion disk survival in rapidly recurring novae.

\end{abstract}


%===================================================================================================
%%% Introduction
%===================================================================================================

\section{Introduction} \label{sec:intro}

Nova eruptions are thermonuclear runaways (TNR) on the surfaces of accreting white dwarfs (WD) \citep{Starr17}. These events occur in interacting binary systems where the companion (donor) star transfers H-rich material to the WD, typically through an accretion disk \citep{Kraft_1964}. As material accumulates on the WD surface, the density and temperature increase until a TNR is triggered, increasing the optical brightness by several orders of magnitude and ejecting the accreted material \citep{Kawash_2021}. Classical novae have only been observed in eruption once and are generally characterised by main-sequence secondaries \citep{Darnley_2012}, with predicted recurrence times ranging from hundreds to millions of years \citep{Bath1978}.

Recurrent novae (RNe) represent a subclass that have been observed to erupt at least twice within a century \citep{Pagnotta_2014}. Systems with recurrence periods $<$10 years are further classified as rapidly recurring novae \citep[RRNe;][]{Darnley_2020}. Sustaining such short recurrence times requires a near-Chandrasekhar-mass ($M_{\text{Ch}}$) WD and a high accretion rate \citep{Yaron_2005,kato14}. The donor stars in RNe are typically more evolved, consisting of subgiants or red giants, and capable of driving elevated mass-transfer rates \citep{gca99,Darnley_2012}. Because the WD in these systems may gain more mass through accretion than it loses during eruptions \citep{Hachisu_2026}, RNe are considered one of the candidates for Type Ia supernovae through the single-degenerate channel \citep{Hernanz_2008,Starrfield_2012a, Dar16a, Hill16}. 

A key question in the study of RNe concerns the fate of the accretion disk during the nova eruption. In classical novae, the disk is often assumed to be destroyed by the expanding ejecta. However, the physical conditions in RRNe, namely, the massive accretion disks and low ejecta masses, may allow a substantial fraction of the disk to survive the eruption \citep{Figueira_2025, Lechuga_2025}. Determining whether the disk survives, and how rapidly accretion resumes after an eruption, is therefore crucial to understand the mass-growth efficiency of these systems.

M31N~2008-12a is the most extreme RRN currently known, exhibiting an approximately annual eruption cycle \citep{tan14}. The details of all the known eruptions are listed in Table~\ref{tab:eruption_all}. Its first recorded eruption occurred in 2008 \citep{nis08}, while archival observations have revealed earlier eruptions in 1992, 1993, and 2001. Since 2013, the system has been monitored extensively across near infrared, optical, ultraviolet, and X-ray wavelengths (2013 -- \citealt{Dar14,Hen14,tan14}; 2014 -- \citealt{dar15,hen15}; 2015 -- \citealt{Dar16a,dar17a,dar17b}; 2016 -- \citealt{Henze_2018}; 2017--2022 -- \citealt{basu_2024}; 2024 -- \citealt{Banerjee_2025a}). The system hosts one of the most massive accreting WDs currently known ($\sim1.38~M_{\odot}$) and exhibits an accretion rate exceeding $10^{-7}~M_{\odot}~\mathrm{yr}^{-1}$ \citep{dar17b,Kato20}.

Previous optical and UV observations have compelling evidence that disk-related emission begins to dominate the system only a couple of weeks after eruption, but could also be as early as 4 days, coinciding with the plateau phase and supersoft source (SSS) $t_{on}$ \citep{dar17b}. The UV--X-ray anti-correlation observed around day 6 \citep{Hen14, basu_2024} is also tentatively associated to signature of a disk with a variable accretion rate. Similar signatures have also been reported in other RRNe, including U~Sco \citep{Ness_2012,GCA_2013} and LMCN~1968-12a \citep{Basu_2025a}. However, most evidence for disk survival has thus far been indirect, relying on photometric or spectroscopic signatures that trace the re-emergence of accretion activity rather than the dynamical properties of the accretion flow itself. 

RRNe can be in close binaries hosting main-sequence or subgiant donors with orbital periods typically of the order of a few hours to $\sim$2\,days. Any disk-related emission modulation would manifest as UV variability on these timescales. This has been demonstrated in RRNe such as U~Sco, with an orbital period of 1.23\,days \citep{Thoroughgood_2001} and LMCN~1968-12a, with a period of 1.26\,days \citep{Kuin_2020}, where UV modulations consistent with the orbital period have been detected during the post-eruption plateau phase \citep{Ness_2012, Basu_2025a}. For M31N~2008-12a, the orbital period has been constrained to lie in the range of 5--23\,days \citep{dar17b}, suggesting that any orbital UV modulation would occur on timescales of days. Additionally, the presence of a surviving or reforming accretion disk could give rise to variability on shorter timescales, well within the orbit-wise cadence of AstroSat/Ultraviolet Imaging Telescope (UVIT) observations ($\sim$97\,min per orbit). These provide the primary motivation for searching for hours-scale UV variability in the high-cadence UVIT light curves of M31N~2008-12a.

In this paper, we present in detail high-cadence AstroSat/UVIT's FUV orbit-wise observations obtained during eruptions between 2019 and 2023 in \S \ref{sec:results}. In \S \ref{sec:multiwave_properties}, we also present optical, UV, and X-ray observations of the 2023--2025 eruptions of M31N~2008-12a. In \S \ref{sec:discussion}, we discuss the implications of the variability, the current understanding of accretion resumption and disk survival in rapidly recurring novae and the importance of mass growth in massive WDs.

%===================================================================================================
%%% Observation
%===================================================================================================

\section{Observations}

\subsection{AstroSat Observations}
\begin{figure*}
    \centering
    \includegraphics[width=\linewidth]{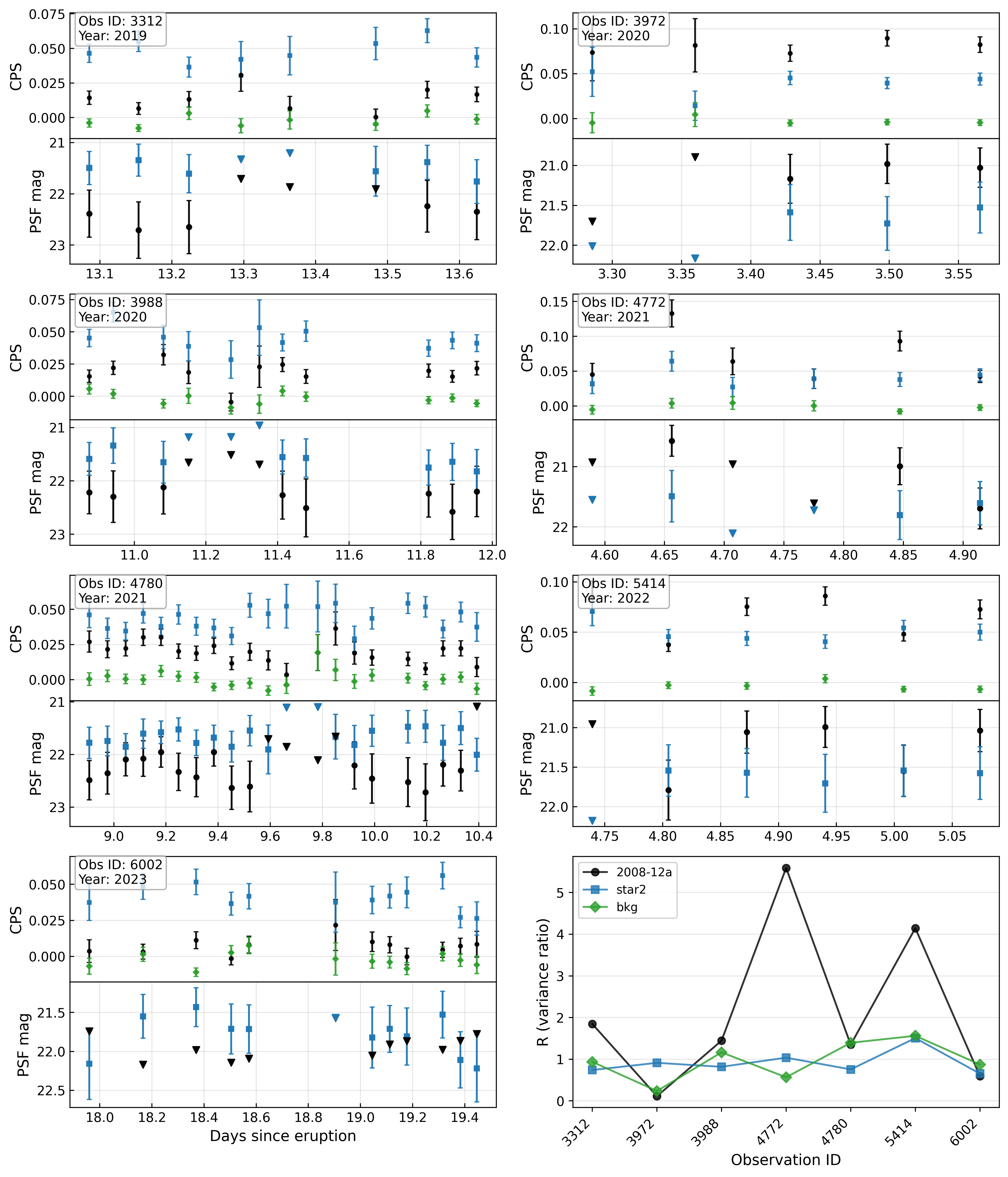}
    \caption{UVIT photometry of M31N 2008-12a \textbf{(black markers)} during its 2019–2023 eruptions. For each panel, the top sub-panel shows \texttt{curvit} aperture photometry and the bottom sub-panel shows PSF photometry, with each point corresponding to an orbit-wise image. Star2 \textbf{(blue markers)} and background \textbf{(green markers)} photometry are overplotted for comparison. The x-axis of each panel gives days since eruption for the corresponding outburst. Left-hand panels correspond to the supersoft source phase, while right-hand panels correspond to the first 3–5 days after eruption. The lower-right panel shows the variability parameter (R) computed for M31N 2008-12a, star2, and the background for each observation, with M31N 2008-12a showing significant variability relative to the comparison star and background in Obs ID 4772 and 5414, which took place 4-5 days after the outburst. The photometric data underlying this figure are available in machine readable format.}
    \label{fig:uv_variability}
\end{figure*}

AstroSat \citep{KPS_2014} is a UV--X-ray space telescope in a low Earth orbit of period $\sim$97\,mins. It has observed the target with UVIT's FUV channel \citep{annapurni16} in F148W filter from 2019 to 2023. UVIT L2 data were downloaded from ISSDC\footnote{\href{https://astrobrowse.issdc.gov.in/astro_archive/archive/Home.jsp}{https://astrobrowse.issdc.gov.in/astro\_archive/archive/Home.jsp}}. The data were analysed using \texttt{curvit} \citep{curvit_2021} for aperture photometry and \texttt{IRAF} \citep{Tody_1993} for PSF photometry. Aperture correction was performed on the PSF photometry using the method described in \cite{basu_2024d} to take care of broad wings in UVIT's PSF. Calibration from counts to flux and AB mag was performed using the relations provided in \cite{tandon17b, Tandon_2020}. The lightcurves obtained from UVIT data are shown in Figure~\ref{fig:uv_variability} and the data behind the figure is available online.

\subsection{Swift Observations}
Publicly available reduced data for the 2023, 2024, and 2025 eruptions from the Neil Gehrels Swift Observatory \citep{Geh04, burrows05, Roming_2005} were downloaded and analysed using \texttt{HEASoft} \citep{heasoft14}. UVOT observed in the $UVW2$ filter, and the data were examined using the \texttt{uvotmaghist} task. An appropriate source region and a background region devoid of any stars were chosen to extract the magnitudes of the object. Photometry of the UVOT mags was calibrated to the AB system \citep{poo08, bre11} and is not corrected for extinction. UVOT's cadence is useful for analysing lightcurve evolution but is not suitable for short-term variability over a few hours. XRT observed the source in 0.3--10\,keV band, but all the photons were observed to be within 0.3--1.5 keV range. The XRT data were analysed using \texttt{XIMAGE sosta/optimize} task to get the count rates corrected for vignetting, dead time loss, background flux, and the PSF of the instrument. The UV data are provided in Table~\ref{tab:Optical_uv_obs} and the X-ray data are given in Table~\ref{tab:xray_obs}.

\subsection{Optical Observations}
Optical photometry was obtained with the 2\,m Himalayan Chandra Telescope (HCT) equipped with the Himalayan Faint Object Spectrograph Camera (HFOSC)\footnote{\href{https://www.iiap.res.in/?q=iao_2m_hfosc}{https://www.iiap.res.in/?q=iao\_2m\_hfosc}} and with the 0.7\,m GROWTH-India Telescope (GIT; \citealt{harsh_2022a}), in the $u'g'r'i'z'$ filters. The HCT data were reduced using standard procedures in \texttt{IRAF} \citep{Tody_1993} while the GIT data were reduced using its automated pipeline. These observations were supplemented by photometric reports from the community submitted to the Astronomer's Telegram, referred to in Table~\ref{tab:eruption_all}.

Optical spectroscopic observations of the 2023, 2024, and 2025 eruptions were obtained with HCT/HFOSC at early phases following each eruption. The spectra were wavelength-calibrated using FeAr arc-lamp exposures and flux-calibrated against the standard star Feige 34 observed on or around the same nights. The photometric data are provided in Table~\ref{tab:Optical_uv_obs} and the spectroscopic observation log is given in Table~\ref{tab:log_spec_obs}.

%===================================================================================================
%%% Analysis
%===================================================================================================

\section{UVIT FUV Analysis}
\label{sec:results}

\subsection{UVIT orbit-wise lightcurves}
\label{sec:lc_extraction}

The UV light curves of M31N~2008-12a and a nearby comparison star (hereafter star2; located at $\alpha$=00:45:28.28, $\delta$=41:54:45.70) were extracted from the UVIT images using the \texttt{curvit} pipeline \citep{curvit_2021}. It performs aperture photometry on individual orbit-wise UVIT frames to produce high-cadence light curves. The background contribution was estimated from a source-free region near the nova and subtracted from the source counts. The resulting light curves are expressed in counts per second (CPS). In addition to CPS light curves, PSF photometry was performed on the same frames to obtain PSF magnitudes using standard \texttt{IRAF} routines \citep{Tody_1993} in \texttt{pyraf} as explained in detail in \cite{basu_2024}. Upper limits were assigned when the source fell below the detection threshold, and PSF fitting did not converge. The CPS light curves and PSF magnitudes for each observation are shown in Figure~\ref{fig:uv_variability}. The PSF photometry and the upper limits are used as consistency checks of the aperture photometry performed by \texttt{curvit}.

\texttt{Curvit} was also used to extract counts in smaller bins of 100\,s to check for variability on a shorter time scale. But the limited S/N shorter bins restricted our analysis to orbit-wise data.

\subsection{FUV Variability Analysis and Possible Periodicity}
\label{sec:variability}

To assess whether M31N~2008-12a is intrinsically variable within each observation of \textit{AstroSat's} visit, we employed a variance-based test following the approach described in \citet{curvit_2021}. For each orbit-wise observation and for each extracted object (M31N~2008-12a, star2, and background), we computed two quantities from the CPS time series
\begin{itemize}
    \item the observed variance of the data set \\
    $V = \frac{1}{N-1} \sum_{i=1}^{N} \left( x_i - \bar{x} \right)^2$
    \item the expected variance due to measurement errors \\
    $S = \frac{1}{N} \sum_{i=1}^{N} \sigma_i^2$ 
\end{itemize}
where $N$ is the number of \textit{AstroSat's} orbits for each data set, $x_i$ is the CPS in $i^{th}$ orbit, $\bar{x}$ is the mean CPS of a data set, and $\sigma_i$ is the error of $x_i$ reported by \texttt{curvit}. 

The ratio $R = V/S$ then provides a dimensionless measure of excess variability intrinsically linked to the source. A value of $R \approx 1$ indicates that the observed scatter is consistent with measurement noise alone, whereas $R \gtrsim 3$ indicates intrinsic variability in excess of what the error bars can account for. These three quantities ($V$, $S$, $R$) are computed independently for M31N~2008-12a, star2, and the background region in each observation. The value of $R$ for each observation and object is shown in the lower-right panel in Figure~\ref{fig:uv_variability}. The comparison star and background are used to rule out any possible variability arising due to the instrument. 

\begin{figure*}
    \centering
    \includegraphics[width=\linewidth]{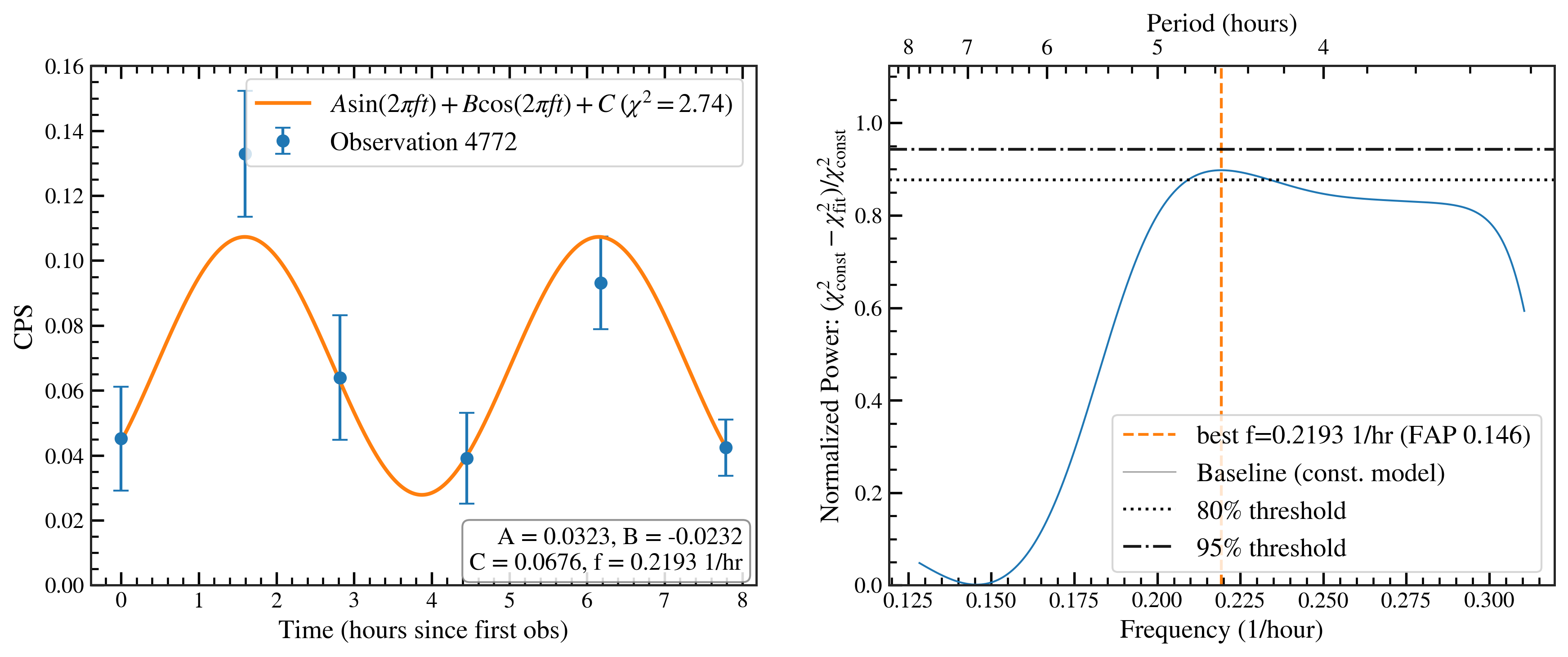}
    \includegraphics[width=\linewidth]{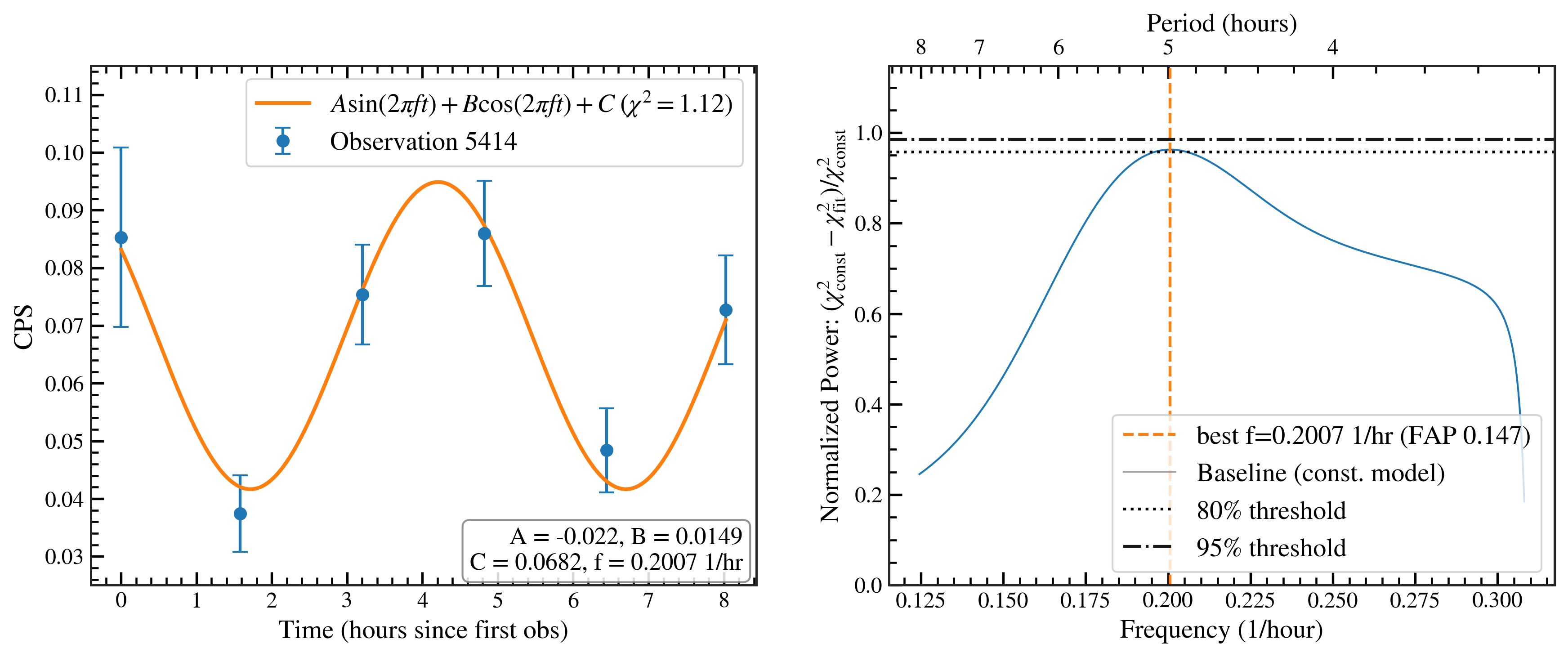}
    \caption{\emph{Left:} Orbit-wise light curves for Observations 4772 (top) and 5414 (bottom), with the best-fit sinusoidal model overplotted. \emph{Right:} The corresponding periodograms, with normalised power plotted against frequency (bottom axis) and period (top axis). The vertical dashed line marks the best-fit frequency and its associated false-alarm probability (FAP), while the horizontal dotted (dot-dashed) line indicates the 80\% (95\%) threshold, i.e. the power level below which 80\% (95\%) of the permuted datasets fall.}
    \label{fig:periodogram}
\end{figure*}

Among the seven observations analysed, significant excess variability of $R > 4$ was found for M31N~2008-12a in two observations: Obs ID 4772 (2021 eruption) and Obs ID 5414 (2022 eruption), shown in Figure~\ref{fig:periodogram}. The comparison star and background remained consistent with $R \sim 1$ in all cases, as shown in Figure~\ref{fig:uv_variability}. This confirms that the variability in these two datasets is astrophysical in origin. It is interesting to note that both these observations were performed during $\sim$4.5-5.1 days after the outburst. 

No variation is found in the Obs ID 3972 (2020 eruption), for which the phase is 3.3-3.6 days after the outburst. It could be because of the low exposure time in the first two orbits of obs 3972, which leads to poor signal-to-noise ratio and higher error bars as seen in Figure~\ref{fig:uv_variability}. The poor statistics constrains our ability to determine the variability. This is further discussed in \S \ref{subsec:SSS_analysis}.

The other four observations where no variability was inferred were all during the supersoft phase (8-20 days after outburst). In this phase, the nova is fainter in FUV by roughly 2 mags than in the initial ($t_0 + 4$~days) phase \citep{basu_2024}. As a result, the errorbars of the orbit-wise images are larger, which puts a limit on the ability to extract the variability. The uncertainties become comparable to the magnitude of variability (V $\sim$ S $\Rightarrow$ R $\sim$ 1). Thus, in the observations during the supersoft phase, even though the number of data points is more, because of poor statistics of each data point, it is not possible to determine the presence of any variability (see \S \ref{subsec:discussion_uv_var}). 

We also searched for periodic signals in the CPS light curves of the nova, the star2 and the background using a weighted least-squares sinusoid fitting method. The method employed is similar to the generalised Lomb-Scargle periodogram of \citet{GLS:2009}, in which the model sinusoid is fit by weighted least-squares at each trial frequency. We evaluate the fit directly on a uniform frequency grid. 
At each trial frequency $f$ we fit the model

\begin{equation}
    m(t) = A\sin(2\pi f t) + B\cos(2\pi f t) + C
\end{equation}

\noindent to the data by weighted linear least squares, with weights $w_i = 1/\sigma_i$ where $\sigma_i$ are the CPS uncertainties. The fit also measures the $\chi^2 (f)$ as a goodness-of-fit.

The periodogram power at each frequency is defined as the improvement over a constant model, normalised to the constant model
 
\begin{equation}
    \mathcal{P}(f) = \frac{\chi^2_\mathrm{const} - \chi^2(f)}{\chi^2_\mathrm{const}},
\end{equation}

\noindent where $\chi^2_\mathrm{const}$ is the $\chi^2$ of the weighted mean model which is a constant. This normalised power ranges from 0 to 1, with values close to unity indicating that the sinusoidal model fits better to the data. The frequency corresponding to the maximum power was then refined using a minimizer in \texttt{scipy} \citep{SciPy20}, and the best-fit coefficients $A$, $B$, and $C$ were computed at this refined frequency.

To assess the statistical significance of the detected peak, we computed the false alarm probability (FAP) using a permutation test. For each permutation, the CPS values and their uncertainties were shuffled along the time axis, as a result, destroying any real periodic signal but preserving the noise properties of the data. The periodogram was then computed on each shuffled dataset, and the maximum power across all trial frequencies was recorded. The FAP was estimated as the fraction of simulated maximum powers that equalled or exceeded the observed peak power.

\begin{figure*}
    \centering
    \includegraphics[width=\linewidth]{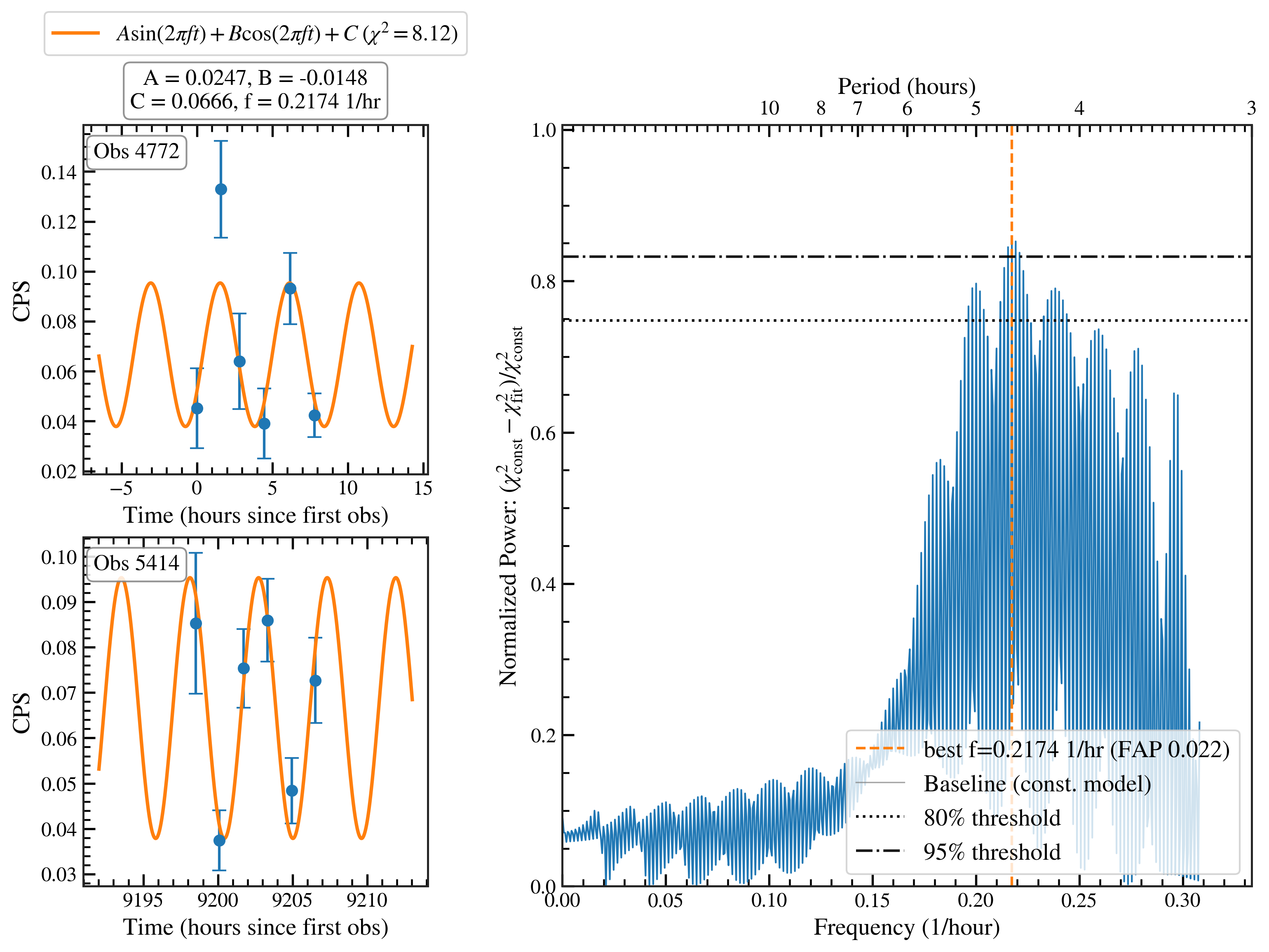}
    \caption{\emph{Left:} The best fit sinusoid for the combined data is overplotted on the observed data. Note that the time axis for both subplots is from the start of Obs 4772. \emph{Right:} Periodogram analysis of the combined data sets (Obs ID 4772 and 5414) which show variability. The best-fit period in the combined dataset is 4.6\,hrs, which is close to the periods in the individual data sets (4.56\,hrs and 4.98\,hrs). The combined periodogram shows a forest of alias peaks arising from the $\sim$383\,day gap between the two observations, a natural consequence of combining two short datasets separated by a long baseline.}
    \label{fig:combined_periodogram}
\end{figure*}

The normalised power $\mathcal{P}(f)$ is used for display purposes in the periodogram, and the FAP thresholds are scaled accordingly before being overplotted in the right side panels of Figure~\ref{fig:periodogram} and \ref{fig:combined_periodogram}.

The permutation approach is preferred over parametric simulations because it makes no assumptions about the noise model, instead relies on the actual distribution of the measured data. Among all the data sets of the nova, star2 and background, none cross the 95\% threshold, commonly used for candidate detection. However, the two datasets which did satisfy the variability criteria show a tentative periodicity detection at $\sim$85\% threshold as shown in Figure~\ref{fig:periodogram}.

The best-fit periods obtained are 4.56 $^{+0.28}_{-0.42}$\,hr and 4.98$^{+0.31}_{-0.31}$\,hr for Observations 4772 and 5414, respectively, with FAP values of 0.146 and 0.147. Neither detection meets the standard 95\% confidence threshold by themselves. 

However, when the two observations, obtained during two different eruptions, are combined into a single dataset, the periodogram yields a best-fit period of $4.60^{+0.25}_{-0.15}$\,hr, consistent with the periods derived from the individual observations (Figure~\ref{fig:combined_periodogram}). The FAP decreases from $\sim$0.15 in each individual observation to 0.02 for the combined data ($\sim$98\% confidence, equivalent to $\sim$2.3$\sigma$). A single sinusoid with a common period provides a statistically acceptable fit to both datasets simultaneously. Although the combined significance is below the conventional 3$\sigma$ threshold, the recovery of a consistent period in two independent eruptions, together with the increase in significance when the data are combined, supports the presence of a candidate $\sim$4.6\,hr modulation with a possible common physical origin in both eruptions.

We note that the best-fit periods from the individual observations are within error bars. The spread may reflect the limited data points available per observation, or genuine period variability between eruptions. Future higher-cadence observations might help resolve this.

The robust result of this analysis is the detection of an intrinsic hours-scale FUV variability in two independent UVIT observations 4-5 days after the outburst. A similar period is also recovered from the periodogram analysis when these two datasets are combined. The recurrence of the inferred 4.5–5\,hrs modulation across multiple epochs during 4-5 days after eruption motivates consideration of a physical origin which is discussed further in \S \ref{sec:discussion}.

\subsection{Analysis of the Supersoft phase}
\label{subsec:SSS_analysis}
To assess whether the absence of variability during the SSS phase is physical in nature or a consequence of reduced sensitivity, we performed two complementary tests. First, we artificially reduced the signal-to-noise ratio (SNR) of the variable-phase observations (Obs~4772 and 5414) to match the mean SNR of the SSS-phase observations by inflating their error bars proportionately. By degrading the signal in these two datasets, the variability parameter drops to $R \ll 3$ for both observations, indicating that the variability would be undetectable at the SSS-phase sensitivity levels. This tells us that the non-detection of variability during the SSS phase can be due to limited sensitivity. The noisy SSS phase data are insufficient to rule out hour-scale variability at $t_0+$ 8-20 days, as seen during the early phase at $t_0 + 4$--5\,days.

However, the SNR argument alone cannot distinguish if a real 4.6\,hr modulation is present in the SSS phase, but hidden by noise, or no such modulation exists during the SSS phase. To distinguish between these, we performed a forced sinusoidal fit to each SSS-phase observation individually. We fixed the period at 4.6\,hr and allowed only the phase and amplitude to vary. If a substantial 4.6\,hr signal were present, the fit would recover a significant amplitude and high normalised power. Instead, we find that the best-fit sinusoid amplitudes in the SSS phase are $\lesssim 0.005$\,CPS, which is roughly an order of magnitude smaller than those recovered in Obs 4772 and 5414 ($\sim$0.03--0.04\,CPS). The normalised power is in the range of 0.12--0.21, compared to $\geq$0.90 in the variable observations. This is indicative of the absence of any periodic signal in the SSS phase data. This suggests that the hour-scale modulation detected at $t_0 + (4-5)$\,days is not present at a comparable level during the SSS phase ($t_0 + (8-20)$\,days). This could indicate that during the supersoft phase, either the source of UV photons evolved or it originates at a different location (see \S \ref{subsec:discussion_uv_var}).

\section{Multiwavelength Lightcurves and Spectroscopy}
\label{sec:multiwave_properties}
\subsection{Optical, UV, X-ray lightcurves}

\begin{figure*}
    \centering
    \includegraphics[width=0.85\linewidth]{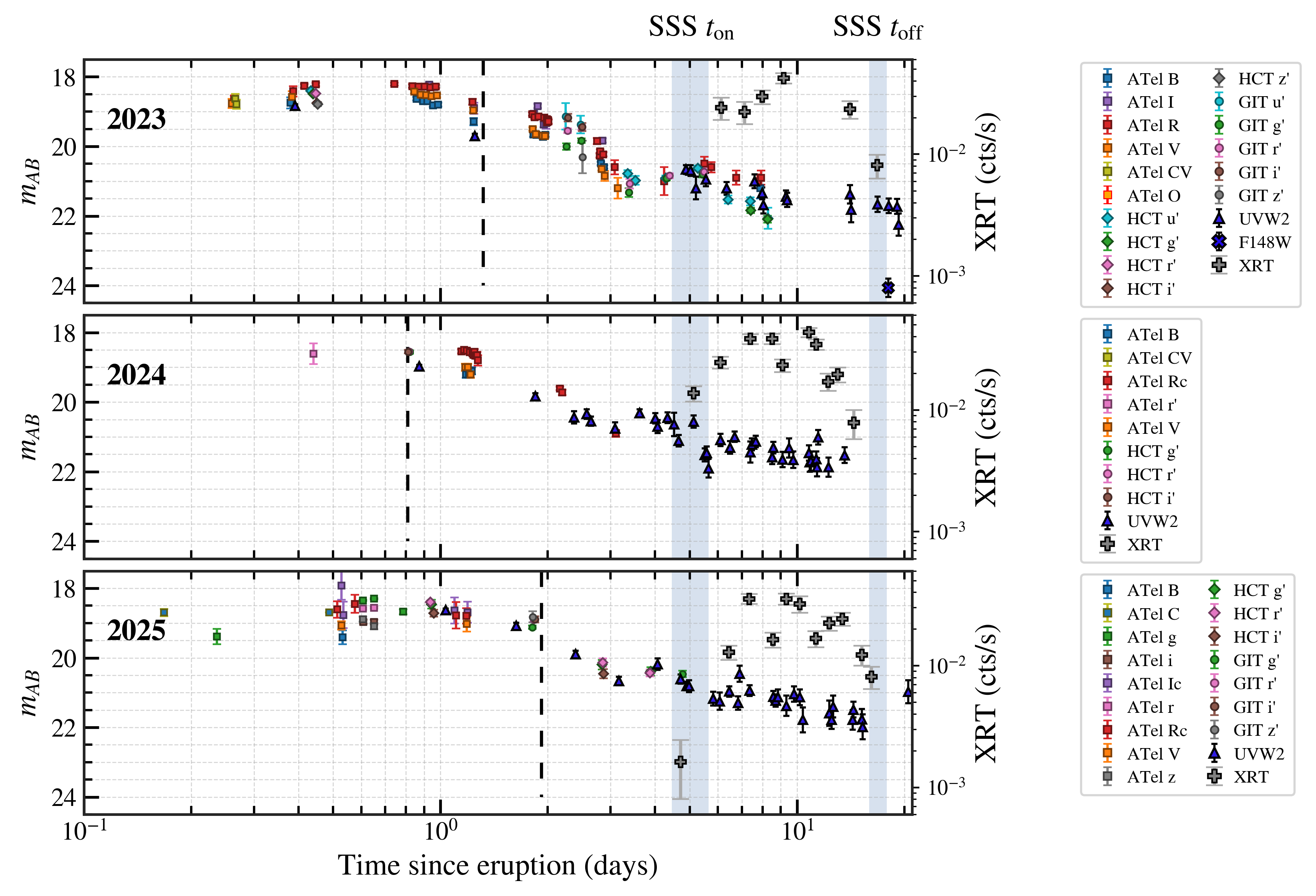}
    \caption{Multiwavelength (optical, UV, X-ray) lightcurves of the 2023-2025 eruptions of M31N~2008-12a. The optical spectroscopy epochs are marked with vertical dashed lines. The supersoft turn-on and turn-off times are shaded in light blue. }
    \label{fig:uvo_lc}
\end{figure*}

\begin{figure*}
    \centering
    \includegraphics[width=0.8\linewidth]{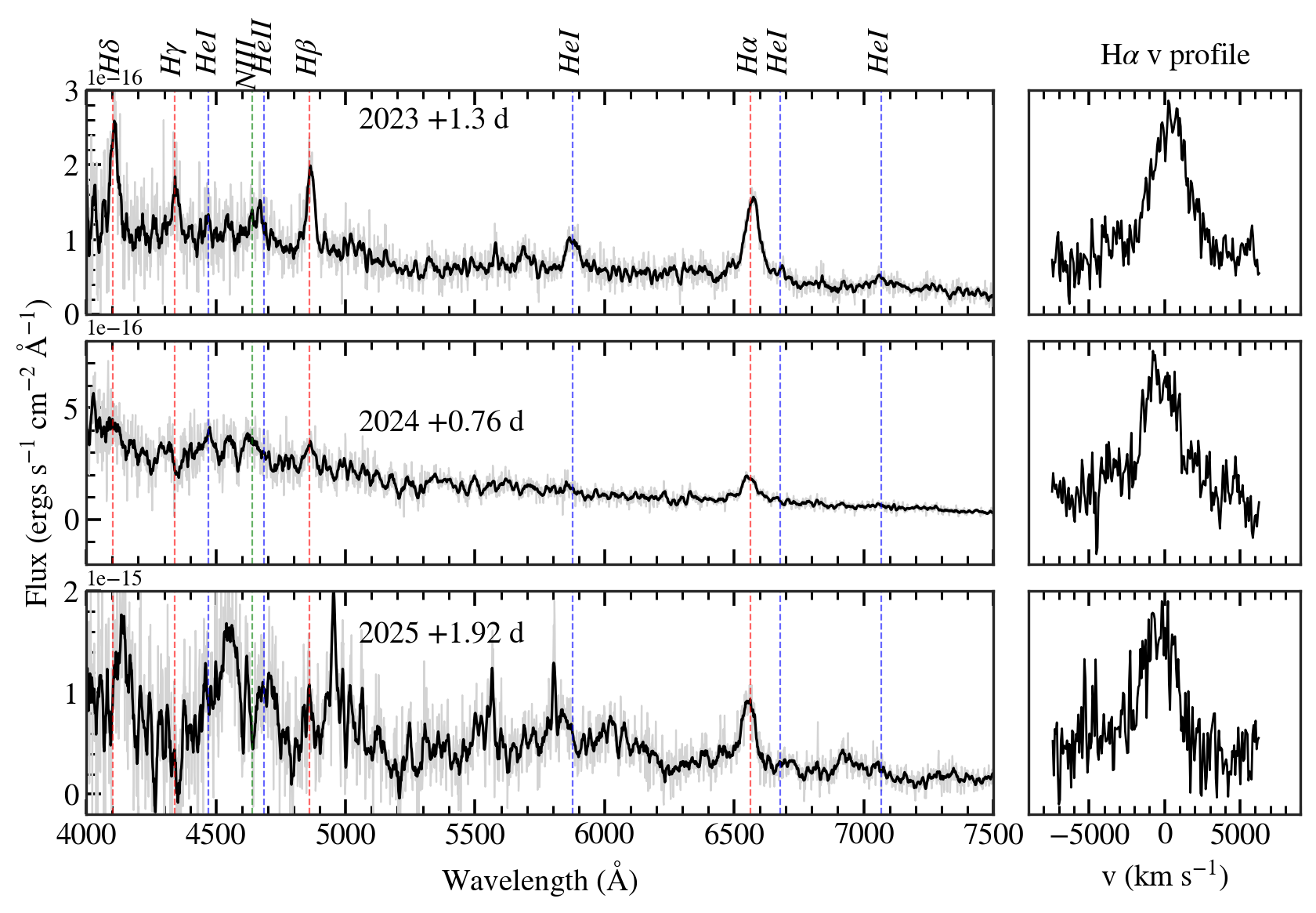}
    \caption{Optical spectra of M31N~2008-12a observed during 2023-2025 outbursts. The panels on the right show the zoomed-in version of the H$\alpha$ emission line and its velocity profiles. }
    \label{fig:spectra}
\end{figure*}

The multiwavelength light curves of 2023, 2024 and 2025 eruptions are shown in Figure~\ref{fig:uvo_lc}. All three eruptions reached a peak optical magnitude of $\sim$18\,mag within the first day, followed by a rapid decline of $\sim$2--3\,mag over the subsequent $\sim$2.5\,days, consistent with the fast decline rate characteristic of this system \citep{Dar14, dar15, Dar16a, Henze_2018, basu_2024}. During the 2023 eruption, the optical light curve exhibits a notable bump or rebrightening immediately after the initial rapid decline around day 4--6. It then resumes its decline toward the plateau phase. A similar feature can not be prominently seen in the 2024 and 2025 light curves, because of the sparser optical coverage in those epochs. 

After the initial steep decline, the $UVW2$ light curve in all three eruptions transitioned into a plateau phase at $\sim$21--22\,mag, persisting until approximately day 15--17. During this plateau, the $UVW2$ light curve exhibits short-term variability on timescales of $\sim$1--2\,days. During the 2023 eruption, the single UVIT/F148W detection at $\sim$day 18 yielded an FUV magnitude of $\sim$24\,mag, significantly fainter than the contemporaneous $UVW2$ measurement.

The Swift/XRT count rate rose steadily from around day 5 ($t_\mathrm{on}$), reaching its peak during the plateau phase, before declining steeply toward the SSS turn-off around day 17--18 ($t_\mathrm{off}$). The X-ray light curve is also markedly variable during the SSS phase, showing flickering on timescales of days. The SSS $t_\mathrm{on}$ and $t_\mathrm{off}$ epochs are marked in each panel of Figure~\ref{fig:uvo_lc} and were adapted from the average values of multiple eruptions studied in \cite{basu_2024}. 

A rebrightening of the UV-optical lightcurves around day 4-6 has been noted in previous eruptions as well \citep{dar15}. The 2023 lightcurve (Figure~\ref{fig:uvo_lc}) confirms the existence of this ``bump'' during the plateau phase. This feature is consistent with the SSS turn-on. \cite{kato15} showed that the early-phase lightcurve dominated by free-free emission fades after 4 days, after which the supersoft emission begins to dominate. Hence, the emerging SSS emission from the WD surface can be reprocessed to produce this bump-like feature.

The overall light curve morphology, including the rapid optical decline, the UV plateau with flickering, and the SSS phase duration, is broadly consistent across all three eruptions (and the previous eruptions). It reinforces the remarkable reproducibility of M31N~2008-12a \citep{Dar16a, basu_2024} from year to year.

\subsection{Optical Spectroscopy}
The spectroscopic epochs correspond to $+1.3$\,d (2023), $+0.76$\,d (2024), and $+1.92$\,d (2025) from the respective eruption dates, and are shown in Figure~\ref{fig:spectra}. All three spectra were obtained during the early phase of the eruption, prior to the SSS $t_\mathrm{on}$. The spectra were dereddened adopting $E(B-V) = 0.1$\,mag \citep{dar17a}, comprising contributions from both Galactic foreground extinction and the internal reddening of M31 along the line of sight to the nova. 

The velocity profiles of the H$\alpha$ emission line extracted from each spectrum are shown in the right panels of Figure~\ref{fig:spectra}. The spectra display broad H Balmer emission lines, He\,\textsc{i} lines at $\lambda\lambda$4471, 5876, 6678, and 7065\,\AA, and a possible He\,\textsc{ii}\,$\lambda$4686, consistent with the He/N spectral classification of this system. The H$\alpha$ velocity profiles reveal a broad, double-peaked or flat-topped structure extending to several thousand km\,s$^{-1}$ on both sides. It indicates a rapidly expanding ejecta with multiple components. The profile morphology is broadly consistent across the three eruptions, as has been reported for earlier eruptions of this system.

Photoionisation modelling using \texttt{CLOUDY} \citep[\texttt{v22.01}]{cloudy17, cloudy23} was performed for the 2023 spectrum, which had the best S/N. The methods applied here are described in \cite{Pav19}. A temperature of $6.31 \times 10^4$\,K and an ionising source luminosity $10^{37}$~erg~s$^{-1}$ were able to obtain the best-fit spectra. A clump ($n_c = 2.5 \times 10^{10}$\,cm$^{-3}$) and a diffuse ($n_d = 2.0 \times 10^9$\,cm$^{-3}$) component in the ratio 7:3 gave a close match with the observed spectrum. The He abundance was tuned to 2.93 He$_{\odot}$ to match the He \textsc{I} $\lambda\lambda$5876\AA~feature. The estimated ejecta mass from this model is $\sim 7.9 \times 10^{-8}$ M$_{\odot}$ \footnote{Ejecta mass derived from density estimates is subject to uncertainties and should be used cautiously (see \citealt{Schaefer_2026})}. These results are consistent with the modelling of the 2018 and 2019 spectra analysed in \cite{basu_2024}.

%===================================================================================================
%%% Discussion
%===================================================================================================
\section{Discussion}
\label{sec:discussion}

\subsection{UV variability on a Few Hours Scale}
\label{subsec:discussion_uv_var}

The most robust result of this work is the detection of intrinsic variability at the timescale of a few hours in the FUV emission of M31N~2008-12a during the immediate post-eruption phase ($t_0$+ 4 to 5\, days). Variability is detected during multiple UVIT visits, demonstrating that the source exhibited short-timescale structure soon after the eruption. Such behaviour is difficult to explain with a low mass ejecta, which has already diffused by day~4 in a fast nova. It instead points to the presence of compact emitting regions.

Previous studies have suggested that the accretion disk in M31N~2008-12a may survive nova eruptions or reform on timescales of 2 weeks \citep{dar17b, basu_2024}. The high-cadence UVIT observations presented here provide a complementary probe by directly tracing variability on very short timescales. The detected FUV variability, therefore, offers a new observational method for investigating the survival and/or re-establishment of accretion structures in RRNe.

Variability is clearly established for two of the three UVIT observations within 5 days of the eruption. The origin of FUV photons during this phase is from a region close to the WD. It can have different physical origins: a residual accretion disk, or accretion streams reforming after the eruption, or material falling back on the WD surface after eruption. Its presence strongly suggests that structures near the WD are already present within days of eruption. This makes M31N~2008-12a an important laboratory for studying how accretion resumes in systems hosting WDs with $M_{\text{WD}}\sim M_{\text{Ch}}$.

Obs ID 3972, despite being within the initial few days of the 2020 outburst, however, does not show variability. While this could be due to poor statistics (see \S \ref{sec:variability}), there could be actual physical reasons too. The phase of this set of observations is between $\sim$3.3-3.6 days, whereas the ones which show variability are both during $\sim$4.5-5.1 days. It could be that the ejecta was still optically thick till day 4, and the disk variability signs are revealed as the ejecta becomes optically thin, from day 4 onwards. Such a scenario is possible in a fast-evolving system like M31N 2008-12a. Variability associated with disk reformation and its associated timescales could also depend on each eruption and may not be similar every year.

The SNR analysis and forced sinusoid fit presented in \S\ref{subsec:SSS_analysis} together suggest that the disappearance of the periodic modulation between the early phase and the SSS phase reflects a physical change in the source rather than a limited sensitivity alone. We argue that this evolution is consistent with the settling down of an initially asymmetric accretion flow into a more azimuthally symmetric disk. At $t_0 + 4$--5\,days, the accretion flow is a clumpy, asymmetric structure whose orbital modulation is detectable in the FUV. The characteristic timescales of the disk at $r_\mathrm{Kep} = 1.56\,R_\odot$ \citep{Frank_2002} are: the orbital period $P_\mathrm{orb} \approx 4.6$\,hr, and the thermal timescale $t_\mathrm{th} \approx 7$\,hr. Over the $\sim$10\,days between the early-phase detection and the SSS phase, $\sim$52 complete orbital periods elapse at $r_\mathrm{Kep}$, which is sufficient for differential rotation to shear away azimuthal asymmetries and allow the disk to settle into a more symmetric configuration. By the SSS phase ($t_0 + 8$--20\,days), the disk may therefore have reached a thermally relaxed, azimuthally symmetric state that produces no net periodic modulation. Another argument for the absence of any periodic modulation is that the UV photons during the supersoft phase could be dominated by those from a different region. An instance of this is the UV emission from near the WD photosphere, which was found to be aperiodic and anti-correlated with the X-ray emission \citep{Hen14,basu_2024}.

\subsection{A Candidate Keplerian Signature close to the WD}

\begin{figure}
    \centering
    \includegraphics[width=\linewidth]{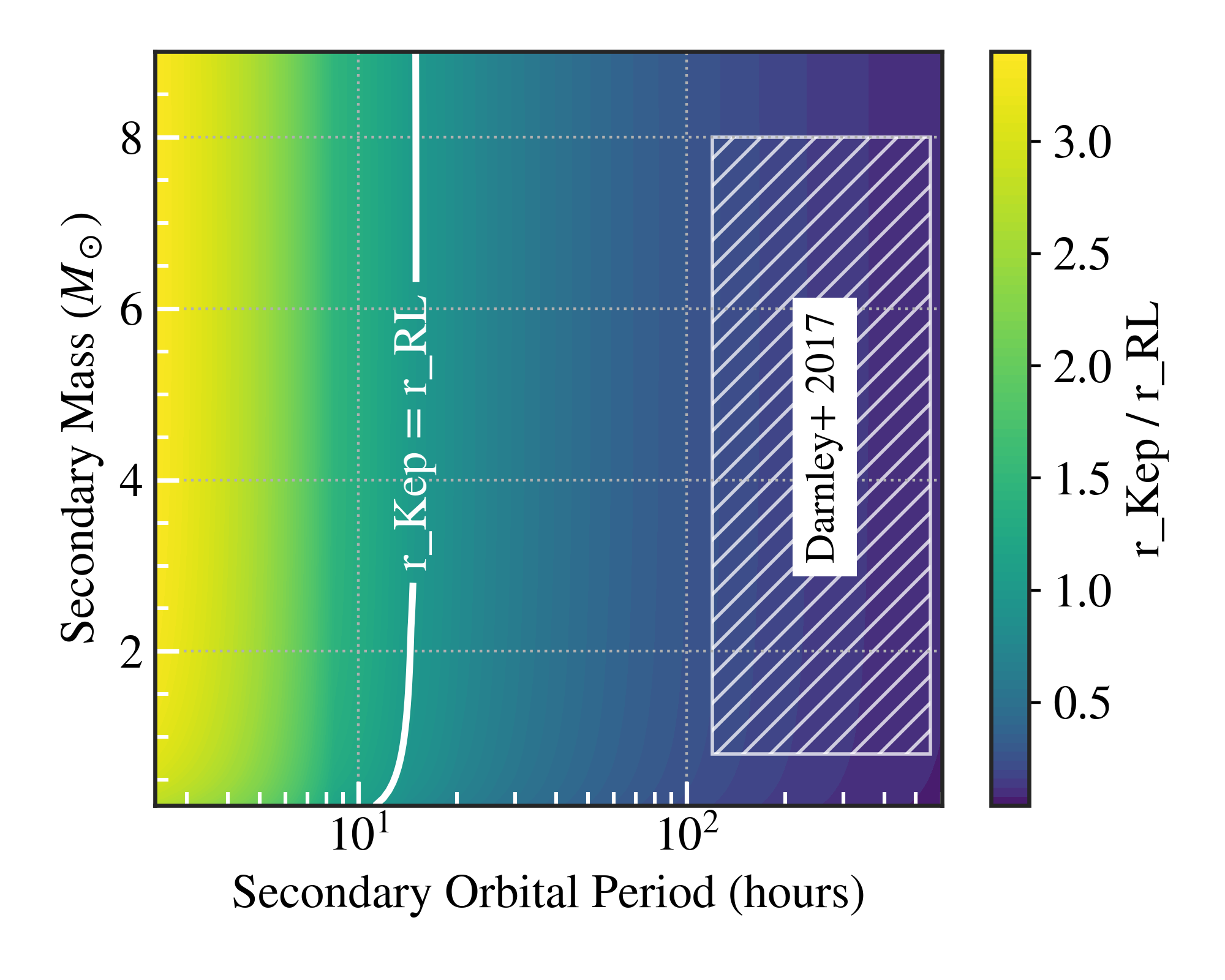}
    \caption{The ratio $r_\mathrm{Kep}/r_\mathrm{RL}$ is shown color-coded as a function of the secondary mass (y-axis) and secondary orbital period (x-axis), for a fixed WD mass $M_1 = 1.38\,M_\odot$. Here, $r_\mathrm{Kep}$ is the radius of a Keplerian orbit around the WD at the tentative 4.6 hr periodicity detected in the UV data, computed from the binary parameters; $r_\mathrm{RL}$ is the Roche lobe radius of the WD. The white contour marks the boundary where $r_\mathrm{Kep}=r_\mathrm{RL}$. The hatched region shows the range of secondary orbital periods and masses constrained for M31N~2008-12a by \cite{dar17b}. Across this entire range, $r_\mathrm{Kep} < r_\mathrm{RL}$, indicating that the source of UV photons lies within the WD's Roche lobe.}
    \label{fig:roche_limit}
\end{figure}

It is intriguing to examine if the 4.6\,hrs period suggested to be present, albeit at only a 2.3 $\sigma$ level, is physically plausible within the known binary configuration of M31N~2008-12a.

To investigate this possibility, we computed the ratio $r_\mathrm{Kep}/r_\mathrm{RL}$ across a grid of secondary masses and orbital periods. Here $r_\mathrm{Kep}$ is the Keplerian radius corresponding to the observed modulation of $\sim$4.6\,hrs and $r_\mathrm{RL}$ is the Roche-lobe radius of the WD estimated using the \citet{Eggleton_1983} approximation. The WD mass was fixed at $M_1 = 1.38\,M_\odot$ \citep{hen15,kato15}.

Figure~\ref{fig:roche_limit} shows all the allowed combinations of donor mass ($0.8 \lesssim M/M_{\odot} \lesssim 8.0$) and donor orbital period ($5 \lesssim P_{\text{orb}} / \text{days} \lesssim 23$) allowed by the constraints in \citet{dar17b} for a Roche lobe filling secondary. The white contour marks the boundary $r_\mathrm{Kep}=r_\mathrm{RL}$, while the hatched region denotes the permitted parameter space of the binary. The inferred Keplerian radius lies comfortably within the Roche lobe of the WD for the system's constrained parameters. 

\begin{figure*}
    \centering
    \includegraphics[width=\linewidth]{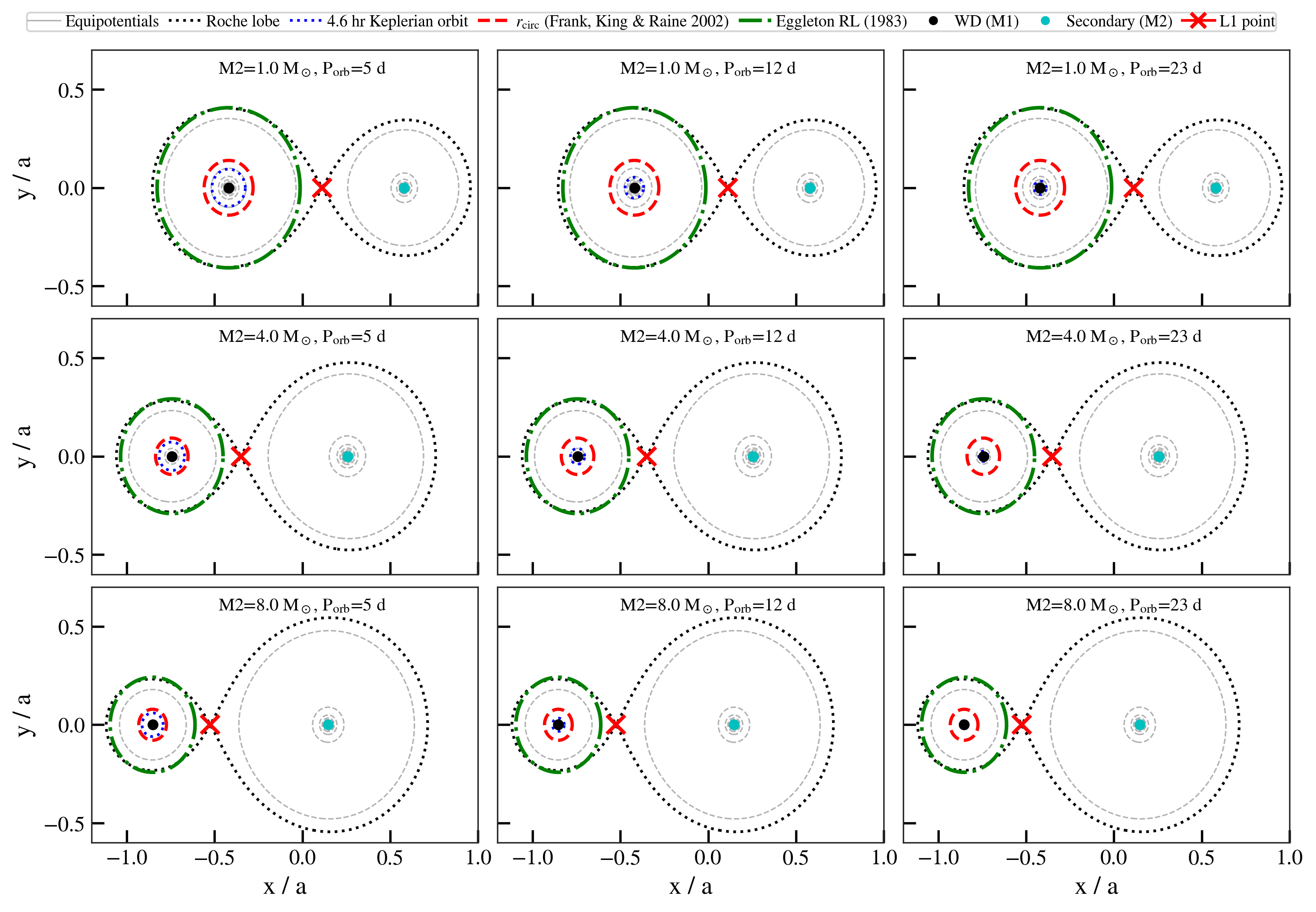}
    \caption{Roche lobe geometry of the M31N~2008-12a binary system for a grid of secondary masses and orbital periods spanning the range constrained by \cite{dar17b}: $M_2=1.0, 4.0, 8.0\,M_\odot$ (rows) and $P_\mathrm{orb} = 5, 12, 23$ days (columns), for a fixed WD mass $M_1 = 1.38\,M_\odot$. In each panel, the equipotential surfaces (dashed grey) and Roche lobes (dotted black, with the \cite{Eggleton_1983} approximation overplotted in green) of both the WD (black) and secondary (cyan) are shown, along with the L1 point (red cross). The tentative Keplerian orbit corresponding to the 4.6\,hr periodicity detected in the UV data is shown around the WD (blue dotted circle), illustrating that this orbit remains well within the WD's Roche lobe across the full range of secondary masses and orbital periods considered. The circularisation radius \citep{Frank_2002} is also shown in a red dashed line.}
    \label{fig:RL_mapping}
\end{figure*}

The Roche-potential geometry shown in Figure~\ref{fig:RL_mapping} further illustrates this result. Across a set of reasonable combinations of donor mass and orbital period, the Keplerian orbit corresponding to a $\sim$4.6\,hr timescale remains well inside the WD Roche lobe. Therefore, if the observed modulation is associated with orbital motion within an accretion flow, the inferred radius is entirely consistent with the known binary parameters of the system.

If the WD instead accretes from the wind of a giant donor, then $P_{\text{orb}} \gg 5$ days \citep{dar17b}. In this case the WD Roche lobe is even larger, and $r_\mathrm{Kep}$ remains well within it (Figure \ref{fig:roche_limit}). The disk is formed by capturing the giant donor's wind, as seen in simulations of RS Oph \citep{Booth_2016}. Its size depends on the wind velocity and orbital speed and separation, and is poorly constrained \citep{val-borro_2009, Pan_2015}.

Any orbital variability should also include a discussion on the inclination of the system. The inclination angle of M31N~2008-12a has been constrained by many previous studies. \cite{dar15} used morphokinematic modelling and suggested the inclination to be $46^{+8}_{-38} {^{\circ}}$, while later studies by \cite{Dar16a, dar17b, dar17a}, \cite{Henze_2018} and \cite{basu_2024}, all suggested a low but non-zero inclination system. Any accretion stream orbiting the hot WD will be irradiated and reprocess the emission in UV wavelengths. At moderate inclinations, partial occultation of this orbiting stream by the WD can produce flux variability consistent with a Keplerian orbital period. The amplitude of the detected modulation is 0.03--0.04 CPS, corresponding to about 40-60\% of the mean flux (see Figure~\ref{fig:periodogram}). This proportion is consistent with what would be expected from a partial occultation at moderate inclination.

The fact that the candidate timescale and variability amplitude correspond to a physically allowed orbit near the WD supports that accretion structures could already be present during the early post-eruption phase.

\subsection{Emerging Population of RRNe with Early Disk Signatures}

M31N~2008-12a is not unique in showing evidence for rapid recovery of the accretion flow after nova outburst. Increasingly, RRNe are exhibiting signatures of disk survival or prompt disk re-establishment shortly after eruption. Examples include U~Sco \citep{Ness_2012,GCA_2013,Muraoka_2025} and LMCN~1968-12a \citep{Basu_2025a}, where observational evidence points to the persistence of accretion-related structures after the eruption. However, in the previously studied systems, disk-related signatures typically emerged on timescales of $\sim$1--2 weeks after outburst. It coincides with the SSS turn-on, when the ejecta sufficiently expand to reveal the central binary. For M31N~2008-12a, evidence of disk signatures were noticed as early as 4 days, coinciding with the onset of the plateau phase and SSS $t_{on}$ \citep{dar17b}. The UV--X-ray anti-correlation detected a week after the eruption \citep{Hen14, basu_2024} is also tentatively linked to the presence of a disk with a changing accretion rate.

The UV variability detected in this study adds a potentially new dimension to this picture. The high-cadence UVIT observations reveal intrinsic FUV variability on timescales of a few hours as early as $\sim$4--5 days after eruption, which is earlier than those reported in other RRNe and comparable to the earliest signatures reported in M31N~2008-12a itself. The candidate modulation discussed here may denote a dynamical tracer of material orbiting within the WD Roche lobe. If confirmed by future observations with larger facilities, such variability could provide a direct means of probing the geometry and evolution of the post-eruption accretion flow.

\subsection{Disk Survival or Rapid Re-establishment?}
The early ($\lesssim$ 5 days of eruption) detection of variability is particularly noteworthy because it suggests that compact accretion structures may already be present while the nova is still in its evolutionary phase. If the variability indeed originates within the accretion flow, it would imply either remarkable disk survival during the eruption or an exceptionally rapid re-establishment of the inner disk. 

To further investigate the physical origin of the $\sim$4.6\,hr modulation, we computed the circularisation radius $r_\mathrm{circ}$, the radius at which material transferred from the donor through the L1 point settles into a circular orbit to conserve angular momentum \citep{Flannery_1975, Lubow_1975}. Following \citet{Frank_2002},

\begin{equation}
    \frac{r_\mathrm{circ}}{a} = \left(1 + \frac{M_2}{M_1}\right)\left(\frac{R_{L1}}{a}\right)^4
    \label{eq:rcirc}
\end{equation}

\noindent where $R_{L1}$ is the distance from the WD to the L1 point. This is the minimum radius at which a disk can form from freshly transferred material. For the stream to reach smaller radii of period 4.6\,hrs, viscous spreading is required, which operates on timescales of $\sim$100 days for a typical viscous disk \citep{Shakura_1973} in this binary system. 

Figure~\ref{fig:RL_mapping} shows $r_\mathrm{Kep}$ and $r_\mathrm{circ}$ overplotted on the Roche potential geometry for a range of secondary masses and orbital periods. Across all combinations explored, $r_\mathrm{Kep} < r_\mathrm{circ}$, meaning the candidate Keplerian structure lies inside the circularisation radius in every case. This implies that newly transferred material from the donor cannot reach $r_\mathrm{Kep}$ directly from accretion alone. Rather, it must first circularise at $r_\mathrm{circ}$ and then spread inward viscously. The modulation is detected $\sim$4--5 days after eruption, well before any viscous spreading \citep[which takes $\sim$100 days;][]{Frank_2002} could deliver material to this radius. The detection of a Keplerian structure within $r_\mathrm{circ}$ at such an early phase is, therefore, consistent with a fraction of the accretion disk surviving the nova eruption.

This represents one of the earliest potential detections of a \textit{surviving} disk-related signature in a recurrent nova and the earliest indication obtained through high-cadence UV variability measurements.

\subsection{Implications for Disk Survival and Sustained Mass Growth}

The growing observational evidence for early accretion signatures in RRNe is broadly consistent with theoretical expectations. Numerical studies suggest that the survival of an accretion disk depends sensitively on the ratio between disk mass and ejecta mass \citep{Lechuga_2025}. Systems containing massive accretion disks and comparatively low ejecta masses are expected to retain a substantial fraction of their pre-eruption accretion structure \citep{Figueira_2025}. 

RRNe naturally satisfy these conditions. Their high accretion rates produce large accretion disks, while the massive WDs responsible for their short recurrence times require only small ignition masses, leading to relatively low ejecta masses \citep{nomo07,kato14,Hill16,Starr17}. Consequently, the eruption becomes incapable of obliterating the accretion disk, allowing accretion to resume almost immediately after outburst. High-resolution 2D simulations by \cite{Lechuga_2025} and low-resolution 3D simulations by \cite{Figueira_2025} showed that their models corresponding to recurrent novae with massive WDs showed signs of disk partially surviving the eruption. 

The results presented here are consistent with this emerging viewpoint. The detection of the FUV variability and the physical plausibility of a Keplerian origin both support the presence of compact accretion structures shortly after eruption. If confirmed through future high-cadence ultraviolet monitoring, such signatures would strengthen the case that RRNe can sustain efficient long-term accretion onto near-Chandrasekhar-mass WDs. This has important implications for understanding WD mass growth towards single-degenerate pathways of Type Ia supernova explosions.

\section{Conclusion}

We have presented high-cadence \textit{AstroSat}/UVIT observations of the recurrent nova M31N~2008-12a obtained during multiple eruptions between 2019 and 2023. The main result is the detection of an intrinsic FUV variability of timescales of a few hours in two of the three observations obtained $\sim$3--5 days after eruption. Differential photometry, followed by variability analysis, confirms that the observed fluctuations are astrophysical in origin rather than instrumental. The four observations performed during the supersoft do not show variability. Although the source is fainter in the FUV during this phase, a forced sinusoidal fit at the 4.6\,hr period yields amplitudes and powers well below those of the early-phase observations. This could occur due to a change in the physical state of the emitting region.

A periodicity search reveals a tentative modulation of $4.5-5$\,hrs in the individual datasets, which is detected with a higher confidence at 4.60 $^{+0.25}_{-0.15}$\,hrs in the combined dataset. We show that a Keplerian orbit corresponding to this timescale lies not only well within the WD's Roche lobe, but also within the circularisation radius around the WD. It suggests that a disk-related or stream-related origin is consistent with the system's known binary parameters. Our findings represent one of the earliest signatures of disk survival in a recurrent nova through high-cadence UV variability measurements.

These results provide a new observational perspective on the question of accretion-disk survival in RRNe. Short-timescale UV variability may be used to probe the compact accretion structures in post-eruption novae. M31N~2008-12a, together with other RRNe, shows evidence for the disk survival and rapid recovery of accretion onto the WD's surface, which is important for the study of mass growth in near $M_{Ch}$ WDs.

\section*{Acknowledgement}
We thank the reviewer for their detailed comments and suggestions, which helped improve the quality of the manuscript. 

We thank the TAC of HCT for allocating time for ToO observations. We thank the staff of IAO, Hanle and CREST, Hosakote, who made these observations possible. The facilities at IAO and CREST are operated by the Indian Institute of Astrophysics, Bangalore. 

We acknowledge the use of GROWTH-India telescope (GIT) data. The GROWTH-India telescope is a 70 cm telescope with a 0.7 deg field of view set up by the Indian Institute of Astrophysics (IIA) and the Indian Institute of Technology Bombay (IITB) with funding from the Indo-US Science and Technology Forum and the Science and Engineering Research Board, Department of Science and Technology, Government of India. It is located at the Indian Astronomical Observatory (IAO, Hanle). We acknowledge funding by the IITB alumni batch of 1994, which partially supports the operation of the telescope. Telescope technical details are available at \href{https://sites.google.com/view/growthindia/}{https://sites.google.com/view/growthindia/}.

This work uses the UVIT data from the AstroSat mission of the Indian Space Research Organisation (ISRO). We thank the AstroSat TAC for allowing us ToO time to observe this nova. We thank the UVIT payload operation centres for verifying and releasing the data via the ISSDC data archive and providing the necessary software tools. 

We acknowledge the use of public data from the {\it{Swift}} data archive. 
This work has also used software and/or web tools obtained from NASA's High Energy Astrophysics Science Archive Research Center (HEASARC), a service of the Goddard Space Flight Center and the Smithsonian Astrophysical Observatory. 

JB thanks Dr Prajwel Joseph for fruitful discussions regarding the use of \texttt{Curvit} software.

GCA and KPS thank the Indian National Science Academy for support under the INSA Senior Scientist Programme.

\facilities{\textit{AstroSat} (UVIT), \textit{Swift} (UVOT and XRT), HCT:2m, GIT:0.7m, }

\software{ %\texttt{CCDLAB} \citep{postma17}, 
            \texttt{PyRAF v2.2.1} \citep{pyraf_2012},
            \texttt{IRAF v2.16.1} \citep{Tody_1993}, 
            \texttt{HEASOFT v6.25}, \texttt{XIMAGE v4.5.1}, and \texttt{XSELECT v2.4e} \citep{heasoft14}, 
            %\texttt{XSPEC v12.12.0} \citep{arnaud96}, 
            \texttt{Python v3.6.6} \citep{python09},  
            \texttt{NumPy} \citep{NumPy20}, 
            \texttt{SciPy} \citep{SciPy20}, 
            \texttt{Pandas} \citep{mckinney10}, 
            \texttt{Matplotlib} \citep{Hunter07}, 
            \texttt{astroquery} \citep{gins19},
            \texttt{astropy} \citep{astropy:2013, astropy:2018, astropy:2022},
            \texttt{Curvit} \citep{curvit_2021}
            }

\appendix

\section{Eruption History}
\begin{table*}[h]
\centering
%\scriptsize
\caption{Record of eruption history of M31N 2008-12a till 2025}

\hspace*{-2.5cm}
\begin{tabular}{llclll}
\toprule
Eruption date$^{(a)}$ & Discovery & Time since last &  SSS $\mathrm{t_{on}}$ date$^{(b)}$ & Detection waveband & Ref. \\
(UT)                  & mag (Filter)      &  eruption (days)    & (UT) & (Observatory) &          \\
\midrule
(1992 Jan 28) & $\cdots$ & $\cdots$ & 1992 Feb 03 & X-ray (ROSAT) & 1, 2 \\
(1993 Jan 03) & $\cdots$ & 341 & 1993 Jan 09 & X-ray (ROSAT) & 1, 2 \\
(2001 Aug 27) & $\cdots$ & $\cdots$ & 2001 Sep 02 & X-ray (Chandra) & 2, 3 \\
2008 Dec 25 & $\cdots$ & $\cdots$ & $\cdots$ & Visible (Miyaki-Argenteus) & 4 \\
2009 Dec 02 & $\cdots$ & 342 & $\cdots$ & Visible (PTF) & 5 \\
2010 Nov 19 & $\cdots$ & 352 & $\cdots$ & Visible (Miyaki-Argenteus) & 2 \\
2011 Oct 22.5 & $\cdots$ & 337.5 & $\cdots$ & Visible (ISON-NM) & 5–8 \\
2012 Oct 18.7 & $\cdots$ & 362.2 & $<$2012 Nov 06.45 & Visible (Miyaki-Argenteus) & 8–11 \\
2013 Nov 26.95 $\pm$ 0.25 & 18.9 (R) & 403.5 & $\le$2013 Dec 03.03 & UV/X-ray (Swift); Visible (iPTF) & 5, 8, 11–14 \\
2014 Oct 02.69 $\pm$ 0.21 & 18.86 ($r^\prime$) & 309.8 $\pm$ 0.7 & 2014 Oct 08.6 $\pm$ 0.5 & UV/X-ray (Swift); Visible (LT) & 8, 15 \\
2015 Aug 28.28 $\pm$ 0.12 & 19.09 ($r^\prime$) & 329.6 $\pm$ 0.3 & 2015 Sep 02.9 $\pm$ 0.7 & UV/X-ray (Swift); Visible (LCO) & 14, 16–18 \\
2016 Dec 12.32 $\pm$ 0.17 & 17.62 (V) & 471.7 $\pm$ 0.2 & 2016 Dec 17.2 $\pm$ 1.1 & UV/X-ray (Swift); Visible (Itagaki) & 19–23 \\
2017 Dec 31.58 $\pm$ 0.20 & 18.41 (clear) & 384.3 $\pm$ 0.4 & 2018 Jan 05.6 $\pm$ 0.5 & UV/X-ray (Swift); Visible (WCO) & 24–27 \\
2018 Nov 06.67 $\pm$ 0.13 & 19.15 & 310.1 $\pm$ 0.3 & 2018 Nov 13.2 $\pm$ 0.5 & UV/X-ray (Swift); Visible (LT) & 28–31 \\
2019 Nov 06.60 $\pm$ 0.11 & 19.40 & 364.9 $\pm$ 0.2 & 2019 Nov 12.4 $\pm$ 0.5 & UV/X-ray (Swift); Visible (HO) & 32–34 \\
2020 Oct 30.49 $\pm$ 0.34 & 18.74 ($g^\prime$) & 358.9 $\pm$ 0.4 & 2020 Nov 05.5 $\pm$ 0.9 & UV/X-ray (Swift); Visible (LT) & 35–38 \\
2021 Nov 14.17 $\pm$ 0.21 & 18.7 (clear) & 379.7 $\pm$ 0.5 & 2021 Nov 19.2 $\pm$ 0.6 & UV/X-ray (Swift); Visible (Itagaki) & 39–42 \\
2022 Dec 02.50 $\pm$ 0.34 & 19.18 ($r^\prime$) & 383.3 $\pm$ 0.5 & 2022 Dec 07.5 $\pm$ 0.5 & UV/X-ray (Swift); Visible (LCOGT) & 43–45 \\
2023 Dec 05.28 $\pm$ 0.07 $^{\dagger}$ & 18.63 (CV) & 367.8 $\pm$ 0.4 & 2023 Dec 11.1 $\pm$ 0.3 & Visible (HMT, XO) & 46–54 \\
2024 Dec 12.99 (discovery) $^{\dagger}$ & 18.56 (CV) & 373.71 & 2024 Dec 18.29 & UV/X-ray (Swift); Visible (BGO SMU) & 55–62 \\
2025 Nov 29.88 $\pm$ 0.24 $^{\dagger}$ & 18.7 (clear) & 351.9 $\pm$ 0.2 & 2025 Dec 04.8 $\pm$ 0.3 & UV/X-ray (Swift); Visible (Itagaki) & 63-69\\ 
\bottomrule
\end{tabular}

\label{tab:eruption_all}

\justifying
$^{\dagger}$ Determined in this work. \\
\textbf{Notes:} Eruption history updated from \citealt{tan14,dar15,hen15a,Dar16a,Henze_2018, basu_2024}. \\
$^{(a)}$ Enclosed in brackets are the archival detections in X-ray data taken from \citealt{hen15a}. \\
$^{(b)}$ ROSAT data was used to estimate the SSS $\rm t_{on}$ for 1992 and 1993. The Chandra detection on 2001 Sep 08 UT was taken as the middle of a typical SSS phase of 12 days to constrain the eruption date. \\
$\cdots$ indicates unavailability of information. \\
References - (1) \cite{whi95}, (2) \cite{hen15a}, (3) \cite{wil04}, (4) \cite{nis08}, (5) \cite{tan14}, (6) \cite{kor11}, (7) \cite{bar11}, (8) \cite{dar15}, (9) \cite{nis12}, (10) \cite{sha12}, (11) \cite{Hen14}, (12) \cite{tan13}, (13) \cite{Dar14}, (14) \cite{Dar16a}, (15) \cite{hen15}, (16) \cite{dar15a}, (17) \cite{dar15b}, (18) \cite{hen15b}, (19) \cite{Darnley_2016b}, (20) \cite{ita16}, (21) \cite{ita16a}, (22) \cite{hen16}, (23) \cite{hen16a}, (24) \cite{boyd17}, (25) \cite{hen18b}, (26) \cite{hen18a}, (27) \cite{nai18},  (28) \cite{hen18c}, (29) \cite{Dar18a}, (30) \cite{Tan18}, (31) \cite{hen18d}, (32) \cite{dar19a}, (33) \cite{Oks19}, (34) \cite{dar19b}, (35) \cite{gal20}, (36) \cite{dar20}, (37) \cite{dar20a}, (38) \cite{dar20b}, (39) \cite{ita21}, (40) \cite{tan21}, (41) \cite{dar21}, (42) \cite{dar21a}, (43) \cite{per22}, (44) \cite{Sha22}, (45) \cite{dar22}, (46) \cite{sun23}, (47) \cite{shafter23}, (48) \cite{perez23}, (49) \cite{basu23}, (50) \cite{balcon23}, (51) \cite{Basu_2023a}, (52) \cite{zhang23}, (53) \cite{shafter_2023}, (54) \cite{Naito_2023}, (55) \cite{Zhao_2024a}, (56) \cite{Page_2024}, (57) \cite{Coca_2024}, (58) \cite{Basu_2024b}, (59) \cite{Shafter_2024b}, (60) \cite{Shafter_2024c}, (61) \cite{Page_2024a}, (62) \cite{Banerjee_2025b}, (63) \cite{Itagaki_2025}, (64) \cite{Basu_2025}, (65) \cite{Diaz_2025}, (66) \cite{Chamoli_2025a}, (67) \cite{Banerjee_2025}, (68) \cite{Darnley_2025}, (69) \cite{Taguchi_2025}.

\end{table*}

\section{Observations and Data}

\label{Appendix_lc}
\begin{table}[]
    \centering
    \caption{Sample of log of optical and UV observations. The full table is available as machine-readable version.}
    \begin{tabular}{cccccc}
    \toprule
    Obs. Date (UTC) & Telescope/Instrument & Filter & Exp Time (s) & AB mag & AB mag error \\
    \midrule
    2023-12-05 17:15:22	& HCT/HFOSC & g$^{\prime}$ & 240  & 18.491 & 0.006 \\
    2023-12-05 17:31:17	& HCT/HFOSC	& i$^{\prime}$ & 120  & 18.753 & 0.026 \\
    2023-12-07 12:37:32	& GIT/Andor	& u$^{\prime}$ & 1000 &	19.138 & 0.382 \\
    2023-12-07 13:17:38	& GIT/Andor & r$^{\prime}$ & 420  & 19.549 & 0.070 \\
    2023-12-05 16:07:01 & Swift/UVOT      & uvw2         & 996.7 & 18.835 & 0.046 \\
    2023-12-06 12:43:08 & Swift/UVOT      & uvw2         & 1012.2 & 19.707 & 0.071 \\
    2023-12-23 07:19:26	& AstroSat/UVIT/FUV & F148W & 11342.77 & 24.065 & 0.265 \\
     \vdots             & \vdots    &\vdots        & \vdots & \vdots & \vdots \\
    \bottomrule
    \end{tabular}
    \label{tab:Optical_uv_obs}
\end{table}

\begin{table}[]
    \centering
    \caption{Sample of log of X-ray (0.3-1.5\,keV) observations. The full table is available as a machine-readable version.}
    \begin{tabular}{ccccc}
    \toprule
    Obs Date (UTC) & Telescope/Instrument & Count Rate & Count rate error & S/N \\
    \midrule
    2023-12-11 09:57:17	& Swift/XRT & 0.024 & 0.005 & 4.87 \\
    2023-12-12 09:41:05 & Swift/XRT & 0.022 & 0.005 & 4.817 \\
    2023-12-13 06:15:31 & Swift/XRT & 0.030 & 0.004 & 7.678 \\
     \vdots             & \vdots    &\vdots        & \vdots & \vdots \\
    \bottomrule
    \end{tabular}
    \label{tab:xray_obs}
\end{table}

\begin{table}[]
    \centering
    \caption{Log of Spectroscopic Observation of M31N 2008-12a during 2023-2025 outbursts.}
    \begin{tabular}{ccccc}
    \toprule
     Proposal ID  & Obs. Date (UTC) & Telescope/Instrument/Grating & Exp Time (s) & Calibration star \\
     \midrule
     HCT-2023-C3-P13 & 2023-12-06 13:58:18 & HCT/HFOSC/Gr7 & 2700 & Feige 34 \\
     HCT-2024-C3-P27 & 2024-12-13 18:01:26 & HCT/HFOSC/Gr7 & 3600 & Feige 34 \\
     HCT-2025-C3-P02 & 2025-12-01 19:10:14 & HCT/HFOSC/Gr7 & 3600 & Feige 34 \\
     \bottomrule
    \end{tabular}

    \label{tab:log_spec_obs}
\end{table}

\newpage
\bibliography{main}{}

@ARTICLE{Pan_2015,
       author = {{Pan}, Kuo-Chuan and {Ricker}, Paul M. and {Taam}, Ronald E.},
        title = "{Simulations of the Symbiotic Recurrent Nova V407 CYG. I. Accretion and Shock Evolutions}",
      journal = {\apj},
         year = 2015,
        month = jun,
       volume = {806},
       number = {1},
          eid = {27},
        pages = {27},
          doi = {10.1088/0004-637X/806/1/27},
archivePrefix = {arXiv},
       eprint = {1503.06181},
 primaryClass = {astro-ph.HE},
       adsurl = {https://ui.adsabs.harvard.edu/abs/2015ApJ...806...27P}
}

@ARTICLE{val-borro_2009,
       author = {{de Val-Borro}, M. and {Karovska}, M. and {Sasselov}, D.},
        title = "{Numerical Simulations of Wind Accretion in Symbiotic Binaries}",
      journal = {\apj},
         year = 2009,
        month = aug,
       volume = {700},
       number = {2},
        pages = {1148-1160},
          doi = {10.1088/0004-637X/700/2/1148},
archivePrefix = {arXiv},
       eprint = {0905.3542},
 primaryClass = {astro-ph.SR},
       adsurl = {https://ui.adsabs.harvard.edu/abs/2009ApJ...700.1148D}
}

@ARTICLE{Booth_2016,
       author = {{Booth}, R.~A. and {Mohamed}, S. and {Podsiadlowski}, Ph.},
        title = "{Modelling the circumstellar medium in RS Ophiuchi and its link to Type Ia supernovae}",
      journal = {\mnras},
         year = 2016,
        month = mar,
       volume = {457},
       number = {1},
        pages = {822-835},
          doi = {10.1093/mnras/stw001},
archivePrefix = {arXiv},
       eprint = {1601.02635},
 primaryClass = {astro-ph.HE},
       adsurl = {https://ui.adsabs.harvard.edu/abs/2016MNRAS.457..822B}
}

@ARTICLE{Thoroughgood_2001,
       author = {{Thoroughgood}, T.~D. and {Dhillon}, V.~S. and {Littlefair}, S.~P. and {Marsh}, T.~R. and {Smith}, D.~A.},
        title = "{The mass of the white dwarf in the recurrent nova U Scorpii}",
      journal = {\mnras},
         year = 2001,
        month = nov,
       volume = {327},
       number = {4},
        pages = {1323-1333},
          doi = {10.1046/j.1365-8711.2001.04828.x},
archivePrefix = {arXiv},
       eprint = {astro-ph/0107477},
 primaryClass = {astro-ph},
       adsurl = {https://ui.adsabs.harvard.edu/abs/2001MNRAS.327.1323T}
}

@ARTICLE{Shakura_1973,
       author = {{Shakura}, N.~I. and {Sunyaev}, R.~A.},
        title = "{Black holes in binary systems. Observational appearance.}",
      journal = {\aap},
         year = 1973,
        month = jan,
       volume = {24},
        pages = {337-355},
       adsurl = {https://ui.adsabs.harvard.edu/abs/1973A&A....24..337S}
}

@BOOK{Frank_2002,
       author = {{Frank}, Juhan and {King}, Andrew and {Raine}, Derek J.},
        title = "{Accretion Power in Astrophysics: Third Edition}",
         year = 2002,
       adsurl = {https://ui.adsabs.harvard.edu/abs/2002apa..book.....F}
}

@ARTICLE{Lubow_1975,
       author = {{Lubow}, S.~H. and {Shu}, F.~H.},
        title = "{Gas dynamics of semidetached binaries.}",
      journal = {\apj},
         year = 1975,
        month = jun,
       volume = {198},
        pages = {383-405},
          doi = {10.1086/153614},
       adsurl = {https://ui.adsabs.harvard.edu/abs/1975ApJ...198..383L}
}

@ARTICLE{Flannery_1975,
       author = {{Flannery}, B.~P.},
        title = "{The location of the hot spot in cataclysmic variable stars as determined from particle trajectories.}",
      journal = {\mnras},
         year = 1975,
        month = feb,
       volume = {170},
        pages = {325-331},
          doi = {10.1093/mnras/170.2.325},
       adsurl = {https://ui.adsabs.harvard.edu/abs/1975MNRAS.170..325F}
}

@software{pyraf_2012,
       author = {{Science Software Branch at STScI}},
        title = "{PyRAF: Python alternative for IRAF}",
 howpublished = {Astrophysics Source Code Library, record ascl:1207.011},
         year = 2012,
        month = jul,
          eid = {ascl:1207.011},
archivePrefix = {ascl},
       eprint = {1207.011},
       adsurl = {https://ui.adsabs.harvard.edu/abs/2012ascl.soft07011S}
}

@INPROCEEDINGS{KPS_2014,
       author = {{Singh}, Kulinder Pal and {Tandon}, S.~N. and {Agrawal}, P.~C. and {Antia}, H.~M. and {Manchanda}, R.~K. and {Yadav}, J.~S. and {Seetha}, S. and {Ramadevi}, M.~C. and {Rao}, A.~R. and {Bhattacharya}, D. and {Paul}, B. and {Sreekumar}, P. and {Bhattacharyya}, S. and {Stewart}, G.~C. and {Hutchings}, J. and {Annapurni}, S.~A. and {Ghosh}, S.~K. and {Murthy}, J. and {Pati}, A. and {Rao}, N.~K. and {Stalin}, C.~S. and {Girish}, V. and {Sankarasubramanian}, K. and {Vadawale}, S. and {Bhalerao}, V.~B. and {Dewangan}, G.~C. and {Dedhia}, D.~K. and {Hingar}, M.~K. and {Katoch}, T.~B. and {Kothare}, A.~T. and {Mirza}, I. and {Mukerjee}, K. and {Shah}, H. and {Shah}, P. and {Mohan}, R. and {Sangal}, A.~K. and {Nagabhusana}, S. and {Sriram}, S. and {Malkar}, J.~P. and {Sreekumar}, S. and {Abbey}, A.~F. and {Hansford}, G.~M. and {Beardmore}, A.~P. and {Sharma}, M.~R. and {Murthy}, S. and {Kulkarni}, R. and {Meena}, G. and {Babu}, V.~C. and {Postma}, J.},
        title = "{ASTROSAT mission}",
    booktitle = {Space Telescopes and Instrumentation 2014: Ultraviolet to Gamma Ray},
         year = 2014,
       editor = {{Takahashi}, Tadayuki and {den Herder}, Jan-Willem A. and {Bautz}, Mark},
       series = {Society of Photo-Optical Instrumentation Engineers (SPIE) Conference Series},
       volume = {9144},
        month = jul,
          eid = {91441S},
        pages = {91441S},
          doi = {10.1117/12.2062667},
       adsurl = {https://ui.adsabs.harvard.edu/abs/2014SPIE.9144E..1SS}
}

@ARTICLE{Tandon_2020,
       author = {{Tandon}, S.~N. and {Postma}, J. and {Joseph}, P. and {Devaraj}, A. and {Subramaniam}, A. and {Barve}, I.~V. and {George}, K. and {Ghosh}, S.~K. and {Girish}, V. and {Hutchings}, J.~B. and {Kamath}, P.~U. and {Kathiravan}, S. and {Kumar}, A. and {Lancelot}, J.~P. and {Leahy}, D. and {Mahesh}, P.~K. and {Mohan}, R. and {Nagabhushana}, S. and {Pati}, A.~K. and {Rao}, N. Kameswara and {Sankarasubramanian}, K. and {Sriram}, S. and {Stalin}, C.~S.},
        title = "{Additional Calibration of the Ultraviolet Imaging Telescope on Board AstroSat}",
      journal = {\aj},
         year = 2020,
        month = apr,
       volume = {159},
       number = {4},
          eid = {158},
        pages = {158},
          doi = {10.3847/1538-3881/ab72a3},
archivePrefix = {arXiv},
       eprint = {2002.01159},
 primaryClass = {astro-ph.IM},
       adsurl = {https://ui.adsabs.harvard.edu/abs/2020AJ....159..158T}
}

@ARTICLE{Darnley_2020,
       author = {{Darnley}, Matthew J. and {Henze}, Martin},
        title = "{On a century of extragalactic novae and the rise of the rapid recurrent novae}",
      journal = {Advances in Space Research},
         year = 2020,
        month = sep,
       volume = {66},
       number = {5},
        pages = {1147-1168},
          doi = {10.1016/j.asr.2019.09.044},
archivePrefix = {arXiv},
       eprint = {1909.10497},
 primaryClass = {astro-ph.SR},
       adsurl = {https://ui.adsabs.harvard.edu/abs/2020AdSpR..66.1147D}
}

@ARTICLE{basu_2024,
       author = {{Basu}, Judhajeet and {Pavana}, M. and {Anupama}, G.~C. and {Barway}, Sudhanshu and {Singh}, Kulinder Pal and {Swain}, Vishwajeet and {Srivastav}, Shubham and {Kumar}, Harsh and {Bhalerao}, Varun and {Sonith}, L.~S. and {Selvakumar}, G.},
        title = "{Multiwavelength Observations of Multiple Eruptions of the Recurrent Nova M31N 2008-12a}",
      journal = {\apj},
         year = 2024,
        month = may,
       volume = {966},
       number = {1},
          eid = {44},
        pages = {44},
          doi = {10.3847/1538-4357/ad2c8e},
archivePrefix = {arXiv},
       eprint = {2310.06586},
 primaryClass = {astro-ph.HE},
       adsurl = {https://ui.adsabs.harvard.edu/abs/2024ApJ...966...44B}
}

@ARTICLE{cloudy17,
       author = {{Ferland}, G.~J. and {Chatzikos}, M. and {Guzm{\'a}n}, F. and {Lykins}, M.~L. and {van Hoof}, P.~A.~M. and {Williams}, R.~J.~R. and {Abel}, N.~P. and {Badnell}, N.~R. and {Keenan}, F.~P. and {Porter}, R.~L. and {Stancil}, P.~C.},
        title = "{The 2017 Release Cloudy}",
      journal = {\rmxaa},
         year = 2017,
        month = oct,
       volume = {53},
        pages = {385-438},
          doi = {10.48550/arXiv.1705.10877},
archivePrefix = {arXiv},
       eprint = {1705.10877},
 primaryClass = {astro-ph.GA},
       adsurl = {https://ui.adsabs.harvard.edu/abs/2017RMxAA..53..385F}
}

@ARTICLE{Pav19,
	author = {{Pavana}, M. and {Anche}, R.~M. and {Anupama}, G.~C. and {Ramaprakash}, A.~N. and 
	{Selvakumar}, G.},
	title = "{Optical spectroscopic and polarization properties of 2011 outburst of the recurrent nova T Pyxidis}",
	journal = {\aap},
	archivePrefix = "arXiv",
	eprint = {1901.00404},
	primaryClass = "astro-ph.SR",
	year = 2019,
	month = feb,
	volume = 622,
	eid = {A126},
	pages = {A126},
	doi = {10.1051/0004-6361/201833728},
	adsurl = {http://adsabs.harvard.edu/abs/2019A%26A...622A.126P}
}

@ARTICLE{cloudy23,
       author = {{Chatzikos}, M. and {Bianchi}, S. and {Camilloni}, F. and {Chakraborty}, P. and {Gunasekera}, C.~M. and {Guzm{\'a}n}, F. and {Milby}, J.~S. and {Sarkar}, A. and {Shaw}, G. and {van Hoof}, P.~A.~M. and {Ferland}, G.~J.},
        title = "{The 2023 Release of Cloudy}",
      journal = {\rmxaa},
         year = 2023,
        month = oct,
       volume = {59},
        pages = {327-343},
          doi = {10.22201/ia.01851101p.2023.59.02.12},
archivePrefix = {arXiv},
       eprint = {2308.06396},
 primaryClass = {astro-ph.GA},
       adsurl = {https://ui.adsabs.harvard.edu/abs/2023RMxAA..59..327C}
}

@ARTICLE{Yaron_2005,
       author = {{Yaron}, O. and {Prialnik}, D. and {Shara}, M.~M. and {Kovetz}, A.},
        title = "{An Extended Grid of Nova Models. II. The Parameter Space of Nova Outbursts}",
      journal = {\apj},
         year = 2005,
        month = apr,
       volume = {623},
       number = {1},
        pages = {398-410},
          doi = {10.1086/428435},
archivePrefix = {arXiv},
       eprint = {astro-ph/0503143},
 primaryClass = {astro-ph},
       adsurl = {https://ui.adsabs.harvard.edu/abs/2005ApJ...623..398Y}
}

@ARTICLE{harsh_2022a,
       author = {{Kumar}, Harsh and {Bhalerao}, Varun and {Anupama}, G.~C. and {Barway}, Sudhanshu and {Basu}, Judhajeet and {Deshmukh}, Kunal and {De}, Kishalay and {Dutta}, Anirban and {Fremling}, Christoffer and {Iyer}, Hrishikesh and {Jassani}, Adeem and {Joharle}, Simran and {Karambelkar}, Viraj and {Khandagale}, Maitreya and {Krishna}, K. Adithya and {Kulkarni}, Sumeet and {Mate}, Sujay and {Patil}, Atharva and {Phanindra}, DVS and {Samantaray}, Subham and {Sharma}, Kritti and {Sharma}, Yashvi and {Shenoy}, Vedant and {Singh}, Avinash and {Srivastava}, Shubham and {Swain}, Vishwajeet and {Waratkar}, Gaurav and {Angchuk}, Dorje and {Dorjay}, Padma and {Dorjai}, Tsewang and {Gyalson}, Tsewang and {Jorphail}, Sonam and {Mahay}, Tashi Thsering and {Norbu}, Rigzin and {Sharma}, Tarun Kumar and {Stanzin}, Jigmet and {Stanzin}, Tsewang and {Stanzin}, Urgain},
        title = "{India's First Robotic Eye for Time-domain Astrophysics: The GROWTH-India Telescope}",
      journal = {\aj},
         year = {2022a},
        month = sep,
       volume = {164},
       number = {3},
          eid = {90},
        pages = {90},
          doi = {10.3847/1538-3881/ac7bea},
archivePrefix = {arXiv},
       eprint = {2206.13535},
 primaryClass = {astro-ph.IM},
       adsurl = {https://ui.adsabs.harvard.edu/abs/2022AJ....164...90K}
}

@ARTICLE{Darnley_2016b,
       author = {{Darnley}, M.~J. and {Rodriguez-Gil}, P. and {Prieto-Arranz}, J. and {et al.}},
        title = "{Recurrent Nova M31N 2008-12a: spectroscopic confirmation of the 2016 eruption}",
      journal = {The Astronomer's Telegram},
         year = 2016,
        month = dec,
       volume = {9852},
        pages = {1},
       adsurl = {https://ui.adsabs.harvard.edu/abs/2016ATel.9852....1D}
}

@ARTICLE{curvit_2021,
       author = {{Joseph}, P. and {Stalin}, C.~S. and {Tandon}, S.~N. and {Ghosh}, S.~K.},
        title = "{Curvit: An open-source Python package to generate light curves from UVIT data}",
      journal = {Journal of Astrophysics and Astronomy},
         year = 2021,
        month = oct,
       volume = {42},
       number = {2},
          eid = {25},
        pages = {25},
          doi = {10.1007/s12036-020-09680-5},
archivePrefix = {arXiv},
       eprint = {2101.06377},
 primaryClass = {astro-ph.IM},
       adsurl = {https://ui.adsabs.harvard.edu/abs/2021JApA...42...25J}
}

@book{python09,
 author = {Van Rossum, Guido and Drake, Fred L.},
 title = {Python 3 Reference Manual},
 year = {2009},
 isbn = {1441412697},
 publisher = {CreateSpace},
 address = {Scotts Valley, CA}
}

@ARTICLE{SciPy20,
  author  = {Virtanen, Pauli and Gommers, Ralf and Oliphant, Travis E. and
            Haberland, Matt and Reddy, Tyler and Cournapeau, David and
            Burovski, Evgeni and Peterson, Pearu and Weckesser, Warren and
            Bright, Jonathan and {van der Walt}, St{\'e}fan J. and
            Brett, Matthew and Wilson, Joshua and Millman, K. Jarrod and
            Mayorov, Nikolay and Nelson, Andrew R. J. and Jones, Eric and
            Kern, Robert and Larson, Eric and Carey, C J and
            Polat, {\.I}lhan and Feng, Yu and Moore, Eric W. and
            {VanderPlas}, Jake and Laxalde, Denis and Perktold, Josef and
            Cimrman, Robert and Henriksen, Ian and Quintero, E. A. and
            Harris, Charles R. and Archibald, Anne M. and
            Ribeiro, Ant{\^o}nio H. and Pedregosa, Fabian and
            {van Mulbregt}, Paul and {SciPy 1.0 Contributors}},
  title   = {{{SciPy} 1.0: Fundamental Algorithms for Scientific
            Computing in Python}},
  journal = {Nature Methods},
  year    = {2020},
  volume  = {17},
  pages   = {261--272},
  adsurl  = {https://rdcu.be/b08Wh},
  doi     = {10.1038/s41592-019-0686-2},
}

@article{astropy:2013,
Adsurl = {http://adsabs.harvard.edu/abs/2013A%26A...558A..33A},
Archiveprefix = {arXiv},
Author = {{Astropy Collaboration} and {Robitaille}, T.~P. and {Tollerud}, E.~J. and {Greenfield}, P. and {Droettboom}, M. and {Bray}, E. and {Aldcroft}, T. and {Davis}, M. and {Ginsburg}, A. and {Price-Whelan}, A.~M. and {Kerzendorf}, W.~E. and {Conley}, A. and {Crighton}, N. and {Barbary}, K. and {Muna}, D. and {Ferguson}, H. and {Grollier}, F. and {Parikh}, M.~M. and {Nair}, P.~H. and {Unther}, H.~M. and {Deil}, C. and {Woillez}, J. and {Conseil}, S. and {Kramer}, R. and {Turner}, J.~E.~H. and {Singer}, L. and {Fox}, R. and {Weaver}, B.~A. and {Zabalza}, V. and {Edwards}, Z.~I. and {Azalee Bostroem}, K. and {Burke}, D.~J. and {Casey}, A.~R. and {Crawford}, S.~M. and {Dencheva}, N. and {Ely}, J. and {Jenness}, T. and {Labrie}, K. and {Lim}, P.~L. and {Pierfederici}, F. and {Pontzen}, A. and {Ptak}, A. and {Refsdal}, B. and {Servillat}, M. and {Streicher}, O.},
Doi = {10.1051/0004-6361/201322068},
Eid = {A33},
Eprint = {1307.6212},
Journal = {\aap},
Month = oct,
Pages = {A33},
Primaryclass = {astro-ph.IM},
Title = {{Astropy: A community Python package for astronomy}},
Volume = 558,
Year = 2013}

@ARTICLE{astropy:2018,
       author = {{Astropy Collaboration} and {Price-Whelan}, A.~M. and
         {Sip{\H{o}}cz}, B.~M. and {G{\"u}nther}, H.~M. and {Lim}, P.~L. and
         {Crawford}, S.~M. and {Conseil}, S. and {Shupe}, D.~L. and
         {Craig}, M.~W. and {Dencheva}, N. and {Ginsburg}, A. and {Vand
        erPlas}, J.~T. and {Bradley}, L.~D. and {P{\'e}rez-Su{\'a}rez}, D. and
         {de Val-Borro}, M. and {Aldcroft}, T.~L. and {Cruz}, K.~L. and
         {Robitaille}, T.~P. and {Tollerud}, E.~J. and {Ardelean}, C. and
         {Babej}, T. and {Bach}, Y.~P. and {Bachetti}, M. and {Bakanov}, A.~V. and
         {Bamford}, S.~P. and {Barentsen}, G. and {Barmby}, P. and
         {Baumbach}, A. and {Berry}, K.~L. and {Biscani}, F. and {Boquien}, M. and
         {Bostroem}, K.~A. and {Bouma}, L.~G. and {Brammer}, G.~B. and
         {Bray}, E.~M. and {Breytenbach}, H. and {Buddelmeijer}, H. and
         {Burke}, D.~J. and {Calderone}, G. and {Cano Rodr{\'\i}guez}, J.~L. and
         {Cara}, M. and {Cardoso}, J.~V.~M. and {Cheedella}, S. and {Copin}, Y. and
         {Corrales}, L. and {Crichton}, D. and {D'Avella}, D. and {Deil}, C. and
         {Depagne}, {\'E}. and {Dietrich}, J.~P. and {Donath}, A. and
         {Droettboom}, M. and {Earl}, N. and {Erben}, T. and {Fabbro}, S. and
         {Ferreira}, L.~A. and {Finethy}, T. and {Fox}, R.~T. and
         {Garrison}, L.~H. and {Gibbons}, S.~L.~J. and {Goldstein}, D.~A. and
         {Gommers}, R. and {Greco}, J.~P. and {Greenfield}, P. and
         {Groener}, A.~M. and {Grollier}, F. and {Hagen}, A. and {Hirst}, P. and
         {Homeier}, D. and {Horton}, A.~J. and {Hosseinzadeh}, G. and {Hu}, L. and
         {Hunkeler}, J.~S. and {Ivezi{\'c}}, {\v{Z}}. and {Jain}, A. and
         {Jenness}, T. and {Kanarek}, G. and {Kendrew}, S. and {Kern}, N.~S. and
         {Kerzendorf}, W.~E. and {Khvalko}, A. and {King}, J. and {Kirkby}, D. and
         {Kulkarni}, A.~M. and {Kumar}, A. and {Lee}, A. and {Lenz}, D. and
         {Littlefair}, S.~P. and {Ma}, Z. and {Macleod}, D.~M. and
         {Mastropietro}, M. and {McCully}, C. and {Montagnac}, S. and
         {Morris}, B.~M. and {Mueller}, M. and {Mumford}, S.~J. and {Muna}, D. and
         {Murphy}, N.~A. and {Nelson}, S. and {Nguyen}, G.~H. and
         {Ninan}, J.~P. and {N{\"o}the}, M. and {Ogaz}, S. and {Oh}, S. and
         {Parejko}, J.~K. and {Parley}, N. and {Pascual}, S. and {Patil}, R. and
         {Patil}, A.~A. and {Plunkett}, A.~L. and {Prochaska}, J.~X. and
         {Rastogi}, T. and {Reddy Janga}, V. and {Sabater}, J. and
         {Sakurikar}, P. and {Seifert}, M. and {Sherbert}, L.~E. and
         {Sherwood-Taylor}, H. and {Shih}, A.~Y. and {Sick}, J. and
         {Silbiger}, M.~T. and {Singanamalla}, S. and {Singer}, L.~P. and
         {Sladen}, P.~H. and {Sooley}, K.~A. and {Sornarajah}, S. and
         {Streicher}, O. and {Teuben}, P. and {Thomas}, S.~W. and
         {Tremblay}, G.~R. and {Turner}, J.~E.~H. and {Terr{\'o}n}, V. and
         {van Kerkwijk}, M.~H. and {de la Vega}, A. and {Watkins}, L.~L. and
         {Weaver}, B.~A. and {Whitmore}, J.~B. and {Woillez}, J. and
         {Zabalza}, V. and {Astropy Contributors}},
        title = "{The Astropy Project: Building an Open-science Project and Status of the v2.0 Core Package}",
      journal = {\aj},
         year = 2018,
        month = sep,
       volume = {156},
       number = {3},
          eid = {123},
        pages = {123},
          doi = {10.3847/1538-3881/aabc4f},
archivePrefix = {arXiv},
       eprint = {1801.02634},
 primaryClass = {astro-ph.IM},
       adsurl = {https://ui.adsabs.harvard.edu/abs/2018AJ....156..123A}
}

@ARTICLE{astropy:2022,
       author = {{Astropy Collaboration} and {Price-Whelan}, Adrian M. and {Lim}, Pey Lian and {Earl}, Nicholas and {Starkman}, Nathaniel and {Bradley}, Larry and {Shupe}, David L. and {Patil}, Aarya A. and {Corrales}, Lia and {Brasseur}, C.~E. and {N{"o}the}, Maximilian and {Donath}, Axel and {Tollerud}, Erik and {Morris}, Brett M. and {Ginsburg}, Adam and {Vaher}, Eero and {Weaver}, Benjamin A. and {Tocknell}, James and {Jamieson}, William and {van Kerkwijk}, Marten H. and {Robitaille}, Thomas P. and {Merry}, Bruce and {Bachetti}, Matteo and {G{"u}nther}, H. Moritz and {Aldcroft}, Thomas L. and {Alvarado-Montes}, Jaime A. and {Archibald}, Anne M. and {B{'o}di}, Attila and {Bapat}, Shreyas and {Barentsen}, Geert and {Baz{'a}n}, Juanjo and {Biswas}, Manish and {Boquien}, M{'e}d{'e}ric and {Burke}, D.~J. and {Cara}, Daria and {Cara}, Mihai and {Conroy}, Kyle E. and {Conseil}, Simon and {Craig}, Matthew W. and {Cross}, Robert M. and {Cruz}, Kelle L. and {D'Eugenio}, Francesco and {Dencheva}, Nadia and {Devillepoix}, Hadrien A.~R. and {Dietrich}, J{"o}rg P. and {Eigenbrot}, Arthur Davis and {Erben}, Thomas and {Ferreira}, Leonardo and {Foreman-Mackey}, Daniel and {Fox}, Ryan and {Freij}, Nabil and {Garg}, Suyog and {Geda}, Robel and {Glattly}, Lauren and {Gondhalekar}, Yash and {Gordon}, Karl D. and {Grant}, David and {Greenfield}, Perry and {Groener}, Austen M. and {Guest}, Steve and {Gurovich}, Sebastian and {Handberg}, Rasmus and {Hart}, Akeem and {Hatfield-Dodds}, Zac and {Homeier}, Derek and {Hosseinzadeh}, Griffin and {Jenness}, Tim and {Jones}, Craig K. and {Joseph}, Prajwel and {Kalmbach}, J. Bryce and {Karamehmetoglu}, Emir and {Ka{l}uszy{'n}ski}, Miko{l}aj and {Kelley}, Michael S.~P. and {Kern}, Nicholas and {Kerzendorf}, Wolfgang E. and {Koch}, Eric W. and {Kulumani}, Shankar and {Lee}, Antony and {Ly}, Chun and {Ma}, Zhiyuan and {MacBride}, Conor and {Maljaars}, Jakob M. and {Muna}, Demitri and {Murphy}, N.~A. and {Norman}, Henrik and {O'Steen}, Richard and {Oman}, Kyle A. and {Pacifici}, Camilla and {Pascual}, Sergio and {Pascual-Granado}, J. and {Patil}, Rohit R. and {Perren}, Gabriel I. and {Pickering}, Timothy E. and {Rastogi}, Tanuj and {Roulston}, Benjamin R. and {Ryan}, Daniel F. and {Rykoff}, Eli S. and {Sabater}, Jose and {Sakurikar}, Parikshit and {Salgado}, Jes{'u}s and {Sanghi}, Aniket and {Saunders}, Nicholas and {Savchenko}, Volodymyr and {Schwardt}, Ludwig and {Seifert-Eckert}, Michael and {Shih}, Albert Y. and {Jain}, Anany Shrey and {Shukla}, Gyanendra and {Sick}, Jonathan and {Simpson}, Chris and {Singanamalla}, Sudheesh and {Singer}, Leo P. and {Singhal}, Jaladh and {Sinha}, Manodeep and {Sip{H{o}}cz}, Brigitta M. and {Spitler}, Lee R. and {Stansby}, David and {Streicher}, Ole and {{{S}}umak}, Jani and {Swinbank}, John D. and {Taranu}, Dan S. and {Tewary}, Nikita and {Tremblay}, Grant R. and {Val-Borro}, Miguel de and {Van Kooten}, Samuel J. and {Vasovi{'c}}, Zlatan and {Verma}, Shresth and {de Miranda Cardoso}, Jos{'e} Vin{'i}cius and {Williams}, Peter K.~G. and {Wilson}, Tom J. and {Winkel}, Benjamin and {Wood-Vasey}, W.~M. and {Xue}, Rui and {Yoachim}, Peter and {Zhang}, Chen and {Zonca}, Andrea and {Astropy Project Contributors}},
        title = "{The Astropy Project: Sustaining and Growing a Community-oriented Open-source Project and the Latest Major Release (v5.0) of the Core Package}",
      journal = {\apj},
         year = 2022,
        month = aug,
       volume = {935},
       number = {2},
          eid = {167},
        pages = {167},
          doi = {10.3847/1538-4357/ac7c74},
archivePrefix = {arXiv},
       eprint = {2206.14220},
 primaryClass = {astro-ph.IM},
       adsurl = {https://ui.adsabs.harvard.edu/abs/2022ApJ...935..167A}
}

@ARTICLE{Shafter_2024b,
       author = {{Shafter}, A.~W.},
        title = "{Follow-up Multicolor Photometry of the Latest Eruption of M31N 2008-12a}",
      journal = {The Astronomer's Telegram},
         year = 2024,
        month = dec,
       volume = {16949},
        pages = {1},
       adsurl = {https://ui.adsabs.harvard.edu/abs/2024ATel16949....1S}
}

@ARTICLE{Page_2024,
       author = {{Page}, K.~L. and {Darnley}, M.~J.},
        title = "{Recurrent nova M31N 2008-12a: Swift/XRT detection of the SSS following the 2024 eruption}",
      journal = {The Astronomer's Telegram},
         year = 2024,
        month = dec,
       volume = {16957},
        pages = {1},
       adsurl = {https://ui.adsabs.harvard.edu/abs/2024ATel16957....1P}
}

@ARTICLE{gca99,
       author = {{Anupama}, G.~C. and {Miko{\l}ajewska}, J.},
        title = "{Recurrent novae at quiescence: systems with giant secondaries}",
      journal = {\aap},
         year = 1999,
        month = apr,
       volume = {344},
        pages = {177-187},
          doi = {10.48550/arXiv.astro-ph/9812432},
archivePrefix = {arXiv},
       eprint = {astro-ph/9812432},
 primaryClass = {astro-ph},
       adsurl = {https://ui.adsabs.harvard.edu/abs/1999A&A...344..177A}
}

@ARTICLE{Shafter_2024c,
       author = {{Shafter}, A.~W.},
        title = "{Deep R-band Photometry of M31N 2008-12a}",
      journal = {The Astronomer's Telegram},
         year = 2024,
        month = dec,
       volume = {16954},
        pages = {1},
       adsurl = {https://ui.adsabs.harvard.edu/abs/2024ATel16954....1S}
}

@ARTICLE{Dar14,
   author = {{Darnley}, M.~J. and {Williams}, S.~C. and {Bode}, M.~F. and 
	{Henze}, M. and {Ness}, J.-U. and {Shafter}, A.~W. and {Hornoch}, K. and 
	{Votruba}, V.},
    title = "{A remarkable recurrent nova in M 31: The optical observations}",
  journal = {\aap},
archivePrefix = "arXiv",
   eprint = {1401.2905},
 primaryClass = "astro-ph.SR",
     year = 2014,
    month = mar,
   volume = 563,
      eid = {L9},
    pages = {L9},
      doi = {10.1051/0004-6361/201423411},
   adsurl = {http://adsabs.harvard.edu/abs/2014A\%26A...563L...9D}
}

@ARTICLE{Hen14,
       author = {{Henze}, M. and {Ness}, J. -U. and {Darnley}, M.~J. and {Bode}, M.~F. and {Williams}, S.~C. and {Shafter}, A.~W. and {Kato}, M. and {Hachisu}, I.},
        title = "{A remarkable recurrent nova in M 31: The X-ray observations}",
      journal = {\aap},
         year = 2014,
        month = mar,
       volume = {563},
          eid = {L8},
        pages = {L8},
          doi = {10.1051/0004-6361/201423410},
archivePrefix = {arXiv},
       eprint = {1401.2904},
 primaryClass = {astro-ph.HE},
       adsurl = {https://ui.adsabs.harvard.edu/abs/2014A&A...563L...8H}
}

@ARTICLE{Page_2024a,
       author = {{Page}, K.~L. and {Darnley}, M.~J.},
        title = "{Swift observations of the 2024 eruption of recurrent nova M31N 2008-12a}",
      journal = {The Astronomer's Telegram},
         year = 2024,
        month = dec,
       volume = {16947},
        pages = {1},
       adsurl = {https://ui.adsabs.harvard.edu/abs/2024ATel16947....1P}
}

@ARTICLE{Banerjee_2025b,
       author = {{Banerjee}, D.~P.~K. and {Geballe}, T.~R. and {Evans}, A. and {Woodward}, C.~E. and {Starrfield}, S.},
        title = "{Near Infrared spectroscopy of the recurrent nova M31N 2008-12a}",
      journal = {The Astronomer's Telegram},
         year = 2025,
        month = jan,
       volume = {16968},
        pages = {1},
       adsurl = {https://ui.adsabs.harvard.edu/abs/2025ATel16968....1B}
}

@ARTICLE{nis08,
   author = {{Nishiyama}, K. and {Kabashima}, F.},
    title = "{Transient Object Followup Reports}",
  journal = {CBAT IAU},
   eprint = {},
     year = 2008,
     month = apr,
     volume = {\href{http://www.cbat.eps.harvard.edu/iau/CBAT_M31.html\#2008-12a}{http://www.cbat.eps.harvard.edu/iau/CBAT\_M31.html\#2008-12a}},
   adsurl = {http://www.cbat.eps.harvard.edu/iau/CBAT_M31.html\#2008-12a}
}

@article{kor11,
  title={CBAT, IAU (http://www.cbat.eps.harvard.edu/unconf/followups)},
  author={{Korotkiy}, S. and {Elenin}, L.},
  journal={J00452885},
  volume={4154094},
  year={2011},
  adsurl = {http://www.cbat.eps.harvard.edu/unconf/followups/J00452885+4154094.html}
}

@article{ita16,
  title={CBAT, IAU (http://www.cbat.eps.harvard.edu/unconf/followups/J00452888+4154097.html)},
  author={{Itagaki}, K.},
  journal={J00452885},
  volume={4154097},
  year={2016},
  adsurl = {http://www.cbat.eps.harvard.edu/unconf/followups/J00452888+4154097.html}
}

@article{nis12,
  title={CBAT, IAU http://www. cbat. eps. harvard. edu/unconf/followups},
  author={{Nishiyama}, K. and {Kabashima}, F.},
  journal={J00452884},
  volume={4154095},
  year={2012},
  adsurl = {http://www.cbat.eps.harvard.edu/unconf/followups/J00452884+4154095.html}
}

@ARTICLE{tan14,
   author = {{Tang}, S. and {Bildsten}, L. and {Wolf}, W.~M. and {Li}, K.~L. and 
	{Kong}, A.~K.~H. and {Cao}, Y. and {Cenko}, S.~B. and {De Cia}, A. and 
	{Kasliwal}, M.~M. and {Kulkarni}, S.~R. and {Laher}, R.~R. and 
	{Masci}, F. and {Nugent}, P.~E. and {Perley}, D.~A. and {Prince}, T.~A. and 
	{Surace}, J.},
    title = "{An Accreting White Dwarf near the Chandrasekhar Limit in the Andromeda Galaxy}",
  journal = {\apj},
archivePrefix = "arXiv",
   eprint = {1401.2426},
 primaryClass = "astro-ph.SR",
     year = 2014,
    month = may,
   volume = 786,
      eid = {61},
    pages = {61},
      doi = {10.1088/0004-637X/786/1/61},
   adsurl = {http://adsabs.harvard.edu/abs/2014ApJ...786...61T}
}

@ARTICLE{dar15,
   author = {{Darnley}, M.~J. and {Henze}, M. and {Steele}, I.~A. and {Bode}, M.~F. and 
	{Ribeiro}, V.~A.~R.~M. and {Rodr{\'{\i}}guez-Gil}, P. and {Shafter}, A.~W. and 
	{Williams}, S.~C. and {Baer}, D. and {Hachisu}, I. and {Hernanz}, M. and 
	{Hornoch}, K. and {Hounsell}, R. and {Kato}, M. and {Kiyota}, S. and 
	{Ku{\v c}{\'a}kov{\'a}}, H. and {Maehara}, H. and {Ness}, J.-U. and 
	{Piascik}, A.~S. and {Sala}, G. and {Skillen}, I. and {Smith}, R.~J. and 
	{Wolf}, M.},
    title = "{A remarkable recurrent nova in M31: Discovery and optical/UV observations of the predicted 2014 eruption}",
  journal = {\aap},
archivePrefix = "arXiv",
   eprint = {1506.04202},
 primaryClass = "astro-ph.SR",
     year = 2015,
    month = aug,
   volume = 580,
      eid = {A45},
    pages = {A45},
      doi = {10.1051/0004-6361/201526027},
   adsurl = {http://adsabs.harvard.edu/abs/2015A\%26A...580A..45D}
}

@ARTICLE{hen15,
   author = {{Henze}, M. and {Ness}, J.-U. and {Darnley}, M.~J. and {Bode}, M.~F. and 
	{Williams}, S.~C. and {Shafter}, A.~W. and {Sala}, G. and {Kato}, M. and 
	{Hachisu}, I. and {Hernanz}, M.},
    title = "{A remarkable recurrent nova in M 31: The predicted 2014 outburst in X-rays with Swift}",
  journal = {\aap},
archivePrefix = "arXiv",
   eprint = {1504.06237},
 primaryClass = "astro-ph.HE",
     year = 2015,
    month = aug,
   volume = 580,
      eid = {A46},
    pages = {A46},
      doi = {10.1051/0004-6361/201526028},
   adsurl = {http://adsabs.harvard.edu/abs/2015A%26A...580A..46H}
}

@ARTICLE{hen15a,
       author = {{Henze}, M. and {Darnley}, M.~J. and {Kabashima}, F. and {Nishiyama}, K. and {Itagaki}, K. and {Gao}, X.},
        title = "{A remarkable recurrent nova in M 31: The 2010 eruption recovered and evidence of a six-month period}",
      journal = {\aap},
         year = 2015,
        month = oct,
       volume = {582},
          eid = {L8},
        pages = {L8},
          doi = {10.1051/0004-6361/201527168},
archivePrefix = {arXiv},
       eprint = {1508.06205},
 primaryClass = {astro-ph.SR},
       adsurl = {https://ui.adsabs.harvard.edu/abs/2015A&A...582L...8H}
}

@ARTICLE{Dar16a,
   author = {{Darnley}, M.~J. and {Henze}, M. and {Bode}, M.~F. and {Hachisu}, I. and 
	{Hernanz}, M. and {Hornoch}, K. and {Hounsell}, R. and {Kato}, M. and 
	{Ness}, J.-U. and {Osborne}, J.~P. and {Page}, K.~L. and {Ribeiro}, V.~A.~R.~M. and 
	{Rodr{\'{\i}}guez-Gil}, P. and {Shafter}, A.~W. and {Shara}, M.~M. and 
	{Steele}, I.~A. and {Williams}, S.~C. and {Arai}, A. and {Arcavi}, I. and 
	{Barsukova}, E.~A. and {Boumis}, P. and {Chen}, T. and {Fabrika}, S. and 
	{Figueira}, J. and {Gao}, X. and {Gehrels}, N. and {Godon}, P. and 
	{Goranskij}, V.~P. and {Harman}, D.~J. and {Hartmann}, D.~H. and 
	{Hosseinzadeh}, G. and {Horst}, J.~C. and {Itagaki}, K. and 
	{Jos{\'e}}, J. and {Kabashima}, F. and {Kaur}, A. and {Kawai}, N. and 
	{Kennea}, J.~A. and {Kiyota}, S. and {Ku{\v c}{\'a}kov{\'a}}, H. and 
	{Lau}, K.~M. and {Maehara}, H. and {Naito}, H. and {Nakajima}, K. and 
	{Nishiyama}, K. and {O'Brien}, T.~J. and {Quimby}, R. and {Sala}, G. and 
	{Sano}, Y. and {Sion}, E.~M. and {Valeev}, A.~F. and {Watanabe}, F. and 
	{Watanabe}, M. and {Williams}, B.~F. and {Xu}, Z.},
    title = "{M31N 2008-12a - The Remarkable Recurrent Nova in M31: Panchromatic Observations of the 2015 Eruption.}",
  journal = {\apj},
archivePrefix = "arXiv",
   eprint = {1607.08082},
 primaryClass = "astro-ph.SR",
     year = 2016,
    month = dec,
   volume = 833,
      eid = {149},
    pages = {149},
      doi = {10.3847/1538-4357/833/2/149},
   adsurl = {http://adsabs.harvard.edu/abs/2016ApJ...833..149D}
}

@ARTICLE{dar17b,
   author = {{Darnley}, M.~J. and {Hounsell}, R. and {Godon}, P. and {Perley}, D.~A. and 
	{Henze}, M. and {Kuin}, N.~P.~M. and {Williams}, B.~F. and {Williams}, S.~C. and 
	{Bode}, M.~F. and {Harman}, D.~J. and {Hornoch}, K. and {Link}, M. and 
	{Ness}, J.-U. and {Ribeiro}, V.~A.~R.~M. and {Sion}, E.~M. and 
	{Shafter}, A.~W. and {Shara}, M.~M.},
    title = "{Inflows, Outflows, and a Giant Donor in the Remarkable Recurrent Nova M31N 2008-12a?{\mdash}Hubble Space Telescope Photometry of the 2015 Eruption}",
  journal = {\apj},
archivePrefix = "arXiv",
   eprint = {1709.10145},
 primaryClass = "astro-ph.HE",
     year = 2017,
    month = nov,
   volume = 849,
      eid = {96},
    pages = {96},
      doi = {10.3847/1538-4357/aa9062},
   adsurl = {http://adsabs.harvard.edu/abs/2017ApJ...849...96D}
}

@ARTICLE{poo08,
   author = {{Poole}, T.~S. and {Breeveld}, A.~A. and {Page}, M.~J. and {Landsman}, W. and 
	{Holland}, S.~T. and {Roming}, P. and {Kuin}, N.~P.~M. and {Brown}, P.~J. and 
	{Gronwall}, C. and {Hunsberger}, S. and {Koch}, S. and {Mason}, K.~O. and 
	{Schady}, P. and {vanden Berk}, D. and {Blustin}, A.~J. and 
	{Boyd}, P. and {Broos}, P. and {Carter}, M. and {Chester}, M.~M. and 
	{Cucchiara}, A. and {Hancock}, B. and {Huckle}, H. and {Immler}, S. and 
	{Ivanushkina}, M. and {Kennedy}, T. and {Marshall}, F. and {Morgan}, A. and 
	{Pandey}, S.~B. and {de Pasquale}, M. and {Smith}, P.~J. and 
	{Still}, M.},
    title = "{Photometric calibration of the Swift ultraviolet/optical telescope}",
  journal = {\mnras},
archivePrefix = "arXiv",
   eprint = {0708.2259},
     year = 2008,
    month = jan,
   volume = 383,
    pages = {627-645},
      doi = {10.1111/j.1365-2966.2007.12563.x},
   adsurl = {http://adsabs.harvard.edu/abs/2008MNRAS.383..627P}
}

@INPROCEEDINGS{bre11,
       author = {{Breeveld}, A.~A. and {Landsman}, W. and {Holland}, S.~T. and {Roming}, P. and {Kuin}, N.~P.~M. and {Page}, M.~J.},
        title = "{An Updated Ultraviolet Calibration for the Swift/UVOT}",
    booktitle = {Gamma Ray Bursts 2010},
         year = 2011,
       editor = {{McEnery}, J.~E. and {Racusin}, J.~L. and {Gehrels}, N.},
       series = {American Institute of Physics Conference Series},
       volume = {1358},
        month = aug,
        pages = {373-376},
          doi = {10.1063/1.3621807},
archivePrefix = {arXiv},
       eprint = {1102.4717},
 primaryClass = {astro-ph.IM},
       adsurl = {https://ui.adsabs.harvard.edu/abs/2011AIPC.1358..373B}
}

@ARTICLE{Tan18,
   author = {{Tan}, H. and {Gao}, X.},
    title = "{Recurrent Nova M31N 2008-12a: Pre-Discovery upper limit and multi-colour follow-up observations with NEXT-XMO}",
  journal = {The Astronomer's Telegram},
     year = 2018,
    month = nov,
   volume = 12200,
   adsurl = {http://adsabs.harvard.edu/abs/2018ATel12200....1T}
}

@ARTICLE{Dar18a,
   author = {{Darnley}, M.~J. and {Henze}, M. and {Shafter}, A.~W. and {Kafka}, S. and 
	{Kato}, M. and {et al.}},
    title = "{Recurrent Nova M31N 2008-12a: discovery of the 2018 eruption}",
  journal = {The Astronomer's Telegram},
     year = 2018,
    month = nov,
   volume = 12177,
   adsurl = {http://adsabs.harvard.edu/abs/2018ATel12177....1D}
}

@ARTICLE{dar17a,
   author = {{Darnley}, M.~J. and {Hounsell}, R. and {Godon}, P. and {Perley}, D.~A. and 
	{Henze}, M. and {Kuin}, N.~P.~M. and {Williams}, B.~F. and {Williams}, S.~C. and 
	{Bode}, M.~F. and {Harman}, D.~J. and {Hornoch}, K. and {Link}, M. and 
	{Ness}, J.-U. and {Ribeiro}, V.~A.~R.~M. and {Sion}, E.~M. and 
	{Shafter}, A.~W. and {Shara}, M.~M.},
    title = "{No Neon, but Jets in the Remarkable Recurrent Nova M31N 2008-12a?{\mdash}Hubble Space Telescope Spectroscopy of the 2015 Eruption}",
  journal = {\apj},
archivePrefix = "arXiv",
   eprint = {1708.06795},
 primaryClass = "astro-ph.SR",
     year = 2017,
    month = sep,
   volume = 847,
      eid = {35},
    pages = {35},
      doi = {10.3847/1538-4357/aa8867},
   adsurl = {http://adsabs.harvard.edu/abs/2017ApJ...847...35D}
}

@ARTICLE{Eggleton_1983,
       author = {{Eggleton}, P.~P.},
        title = "{Aproximations to the radii of Roche lobes.}",
      journal = {\apj},
         year = 1983,
        month = may,
       volume = {268},
        pages = {368-369},
          doi = {10.1086/160960},
       adsurl = {https://ui.adsabs.harvard.edu/abs/1983ApJ...268..368E}
}

@ARTICLE{GLS:2009,
       author = {{Zechmeister}, M. and {K{\"u}rster}, M.},
        title = "{The generalised Lomb-Scargle periodogram. A new formalism for the floating-mean and Keplerian periodograms}",
      journal = {\aap},
         year = 2009,
        month = mar,
       volume = {496},
       number = {2},
        pages = {577-584},
          doi = {10.1051/0004-6361:200811296},
archivePrefix = {arXiv},
       eprint = {0901.2573},
 primaryClass = {astro-ph.IM},
       adsurl = {https://ui.adsabs.harvard.edu/abs/2009A&A...496..577Z}
}

@ARTICLE{Schaefer_2026,
       author = {{Schaefer}, Bradley E.},
        title = "{Orbital Period Changes of Recurrent Nova T Pyxidis Demonstrate that M$_{ejecta}$ {\ensuremath{\gg}} 11.3 {\texttimes} M$_{accreted}$ and that It Is Not a Type Ia Supernova Progenitor}",
      journal = {\apj},
         year = 2026,
        month = apr,
       volume = {1000},
       number = {2},
          eid = {219},
        pages = {219},
          doi = {10.3847/1538-4357/ae3aa2},
archivePrefix = {arXiv},
       eprint = {2602.09377},
 primaryClass = {astro-ph.SR},
       adsurl = {https://ui.adsabs.harvard.edu/abs/2026ApJ..1000..219S}
}

@ARTICLE{Geh04,
       author = {{Gehrels}, N. and {Chincarini}, G. and {Giommi}, P. and {Mason}, K.~O. and {Nousek}, J.~A. and {Wells}, A.~A. and {White}, N.~E. and {Barthelmy}, S.~D. and {Burrows}, D.~N. and {Cominsky}, L.~R. and {Hurley}, K.~C. and {Marshall}, F.~E. and {M{\'e}sz{\'a}ros}, P. and {Roming}, P.~W.~A. and {Angelini}, L. and {Barbier}, L.~M. and {Belloni}, T. and {Campana}, S. and {Caraveo}, P.~A. and {Chester}, M.~M. and {Citterio}, O. and {Cline}, T.~L. and {Cropper}, M.~S. and {Cummings}, J.~R. and {Dean}, A.~J. and {Feigelson}, E.~D. and {Fenimore}, E.~E. and {Frail}, D.~A. and {Fruchter}, A.~S. and {Garmire}, G.~P. and {Gendreau}, K. and {Ghisellini}, G. and {Greiner}, J. and {Hill}, J.~E. and {Hunsberger}, S.~D. and {Krimm}, H.~A. and {Kulkarni}, S.~R. and {Kumar}, P. and {Lebrun}, F. and {Lloyd-Ronning}, N.~M. and {Markwardt}, C.~B. and {Mattson}, B.~J. and {Mushotzky}, R.~F. and {Norris}, J.~P. and {Osborne}, J. and {Paczynski}, B. and {Palmer}, D.~M. and {Park}, H. -S. and {Parsons}, A.~M. and {Paul}, J. and {Rees}, M.~J. and {Reynolds}, C.~S. and {Rhoads}, J.~E. and {Sasseen}, T.~P. and {Schaefer}, B.~E. and {Short}, A.~T. and {Smale}, A.~P. and {Smith}, I.~A. and {Stella}, L. and {Tagliaferri}, G. and {Takahashi}, T. and {Tashiro}, M. and {Townsley}, L.~K. and {Tueller}, J. and {Turner}, M.~J.~L. and {Vietri}, M. and {Voges}, W. and {Ward}, M.~J. and {Willingale}, R. and {Zerbi}, F.~M. and {Zhang}, W.~W.},
        title = "{The Swift Gamma-Ray Burst Mission}",
      journal = {\apj},
         year = 2004,
        month = aug,
       volume = {611},
       number = {2},
        pages = {1005-1020},
          doi = {10.1086/422091},
archivePrefix = {arXiv},
       eprint = {astro-ph/0405233},
 primaryClass = {astro-ph},
       adsurl = {https://ui.adsabs.harvard.edu/abs/2004ApJ...611.1005G}
}

@ARTICLE{gins19,
       author = {{Ginsburg}, Adam and {Sip{\H{o}}cz}, Brigitta M. and {Brasseur}, C.~E. and {Cowperthwaite}, Philip S. and {Craig}, Matthew W. and {Deil}, Christoph and {Guillochon}, James and {Guzman}, Giannina and {Liedtke}, Simon and {Lian Lim}, Pey and {Lockhart}, Kelly E. and {Mommert}, Michael and {Morris}, Brett M. and {Norman}, Henrik and {Parikh}, Madhura and {Persson}, Magnus V. and {Robitaille}, Thomas P. and {Segovia}, Juan-Carlos and {Singer}, Leo P. and {Tollerud}, Erik J. and {de Val-Borro}, Miguel and {Valtchanov}, Ivan and {Woillez}, Julien and {Astroquery Collaboration} and {a subset of astropy Collaboration}},
        title = "{astroquery: An Astronomical Web-querying Package in Python}",
      journal = {\aj},
         year = 2019,
        month = mar,
       volume = {157},
       number = {3},
          eid = {98},
        pages = {98},
          doi = {10.3847/1538-3881/aafc33},
archivePrefix = {arXiv},
       eprint = {1901.04520},
 primaryClass = {astro-ph.IM},
       adsurl = {https://ui.adsabs.harvard.edu/abs/2019AJ....157...98G}
}

@INPROCEEDINGS{annapurni16,
       author = {{Subramaniam}, Annapurni and {Tandon}, Shyam N. and {Hutchings}, John and {Ghosh}, Swarna K. and {George}, Koshy and {Girish}, V. and {Kamath}, P.~U. and {Kathiravan}, S. and {Kumar}, Amit and {Lancelot}, J. Paul and {Mahesh}, P.~K. and {Mohan}, Rekhesh and {Murthy}, Jayant and {Nagabhushana}, S. and {Pati}, Ashok K. and {Postma}, Joe and {Rao}, N. Kameswara and {Sankarasubramanian}, Kasiviswanathan and {Sreekumar}, P. and {Sriram}, S. and {Stalin}, Chelliah S. and {Sutaria}, Firoza and {Sreedhar}, Yuvraj Harsha and {Barve}, Indrajit V. and {Mondal}, Chayan and {Sahu}, Snehalate},
        title = "{In-orbit performance of UVIT on ASTROSAT}",
    booktitle = {Space Telescopes and Instrumentation 2016: Ultraviolet to Gamma Ray},
         year = 2016,
       editor = {{den Herder}, Jan-Willem A. and {Takahashi}, Tadayuki and {Bautz}, Marshall},
       series = {Society of Photo-Optical Instrumentation Engineers (SPIE) Conference Series},
       volume = {9905},
        month = jul,
          eid = {99051F},
        pages = {99051F},
          doi = {10.1117/12.2235271},
archivePrefix = {arXiv},
       eprint = {1608.01073},
 primaryClass = {astro-ph.IM},
       adsurl = {https://ui.adsabs.harvard.edu/abs/2016SPIE.9905E..1FS}
}

@ARTICLE{tandon17b,
       author = {{Tandon}, S.~N. and {Subramaniam}, Annapurni and {Girish}, V. and {Postma}, J. and {Sankarasubramanian}, K. and {Sriram}, S. and {Stalin}, C.~S. and {Mondal}, C. and {Sahu}, S. and {Joseph}, P. and {Hutchings}, J. and {Ghosh}, S.~K. and {Barve}, I.~V. and {George}, K. and {Kamath}, P.~U. and {Kathiravan}, S. and {Kumar}, A. and {Lancelot}, J.~P. and {Leahy}, D. and {Mahesh}, P.~K. and {Mohan}, R. and {Nagabhushana}, S. and {Pati}, A.~K. and {Kameswara Rao}, N. and {Sreedhar}, Y.~H. and {Sreekumar}, P.},
        title = "{In-orbit Calibrations of the Ultraviolet Imaging Telescope}",
      journal = {\aj},
         year = 2017,
        month = sep,
       volume = {154},
       number = {3},
          eid = {128},
        pages = {128},
          doi = {10.3847/1538-3881/aa8451},
archivePrefix = {arXiv},
       eprint = {1705.03715},
 primaryClass = {astro-ph.IM},
       adsurl = {https://ui.adsabs.harvard.edu/abs/2017AJ....154..128T}
}

@ARTICLE{burrows05,
       author = {{Burrows}, David N. and {Hill}, J.~E. and {Nousek}, J.~A. and {Kennea}, J.~A. and {Wells}, A. and {Osborne}, J.~P. and {Abbey}, A.~F. and {Beardmore}, A. and {Mukerjee}, K. and {Short}, A.~D.~T. and {Chincarini}, G. and {Campana}, S. and {Citterio}, O. and {Moretti}, A. and {Pagani}, C. and {Tagliaferri}, G. and {Giommi}, P. and {Capalbi}, M. and {Tamburelli}, F. and {Angelini}, L. and {Cusumano}, G. and {Br{\"a}uninger}, H.~W. and {Burkert}, W. and {Hartner}, G.~D.},
        title = "{The Swift X-Ray Telescope}",
      journal = {\ssr},
         year = 2005,
        month = oct,
       volume = {120},
       number = {3-4},
        pages = {165-195},
          doi = {10.1007/s11214-005-5097-2},
archivePrefix = {arXiv},
       eprint = {astro-ph/0508071},
 primaryClass = {astro-ph},
       adsurl = {https://ui.adsabs.harvard.edu/abs/2005SSRv..120..165B}
}

@ARTICLE{boyd17,
       author = {{Boyd}, D. and {Hornoch}, K. and {Henze}, M. and {Darnley}, M.~J. and {Shafter}, A.~W. and {Kafka}, S. and {Kato}, M. and {et al.}},
        title = "{Recurrent Nova M31N 2008-12a: discovery of the 2017 eruption}",
      journal = {The Astronomer's Telegram},
         year = 2017,
        month = dec,
       volume = {11116},
        pages = {1},
       adsurl = {https://ui.adsabs.harvard.edu/abs/2017ATel11116....1B}
}

@ARTICLE{nai18,
       author = {{Naito}, H. and {Watanabe}, F. and {Sano}, Y. and {Kuramoto}, K. and {Itagaki}, K. and {Kiyota}, S. and {Arai}, A. and {Maehara}, H. and {Matsumoto}, K. and {Fukui}, A. and {Nishiyama}, K. and {Kabashima}, F. and {Henze}, M. and {Darnley}, M.~J. and {Shafter}, A.~W. and {Kato}, M. and {et al.}},
        title = "{Optical Photometric Observations of M31N 2008-12a: Pre- and Post-maximum of the 2017 Eruption}",
      journal = {The Astronomer's Telegram},
         year = 2018,
        month = jan,
       volume = {11133},
        pages = {1},
       adsurl = {https://ui.adsabs.harvard.edu/abs/2018ATel11133....1N}
}

@ARTICLE{Oks19,
       author = {{Oksanen}, A. and {Darnley}, M.~J. and {Shafter}, A.~W. and {Kafka}, S. and {Kato}, M. and {Henze}, M.},
        title = "{Recurrent Nova M31N 2008-12a: discovery of the 2019 eruption}",
      journal = {The Astronomer's Telegram},
         year = 2019,
        month = nov,
       volume = {13269},
        pages = {1},
       adsurl = {https://ui.adsabs.harvard.edu/abs/2019ATel13269....1O}
}

@ARTICLE{dar20,
       author = {{Darnley}, M.~J. and {Shafter}, A.~W. and {Kafka}, S. and {Williams}, S. and {Henze}, M.},
        title = "{Recurrent Nova M31N 2008-12a: discovery of the 2020 eruption}",
      journal = {The Astronomer's Telegram},
         year = 2020,
        month = oct,
       volume = {14130},
        pages = {1},
       adsurl = {https://ui.adsabs.harvard.edu/abs/2020ATel14130....1D}
}

@ARTICLE{ita21,
       author = {{Itagaki}, K. and {Vanmunster}, T. and {Watanabe}, F. and {Carballo}, J. -L.~G. and {Matsumoto}, K. and {Naito}, H. and {Healy}, M.~W. and {Darnley}, M.~J. and {Williams}, S.~C. and {Kafka}, S. and {Shafter}, A.~W. and {Henze}, M.},
        title = "{Recurrent Nova M31N 2008-12a: discovery of the 2021 eruption}",
      journal = {The Astronomer's Telegram},
         year = 2021,
        month = nov,
       volume = {15034},
        pages = {1},
       adsurl = {https://ui.adsabs.harvard.edu/abs/2021ATel15034....1I}
}

@ARTICLE{tan21,
       author = {{Tan}, Hanjie and {Zhang}, Mi and {Zhao}, Jingyuan and {Ruan}, Jiangao and {Sun}, Guoyou and {Gao}, Xing},
        title = "{Recurrent Nova M31N 2008-12a: Pre-detection upper limit and multicolor follow-up observations of 2021 eruption with XMO-NEXT}",
      journal = {The Astronomer's Telegram},
         year = 2021,
        month = nov,
       volume = {15037},
        pages = {1},
       adsurl = {https://ui.adsabs.harvard.edu/abs/2021ATel15037....1T}
}

@ARTICLE{per22,
       author = {{Perez-Fournon}, I. and {Poidevin}, F. and {Aznar Menargues}, G. and {Calatayud-Borras}, Y. and {Casasbuenas}, A. and {Cervera Cortes}, M. and {Cortes-Perez}, I. and {del Socorro Prieto}, J. and {Delgado Mancheno}, M. and {Burjachs}, A. Ferrer I. and {Flores le Roux}, R. and {Galan}, D. and {Geraldia Gonzalez}, S. and {Grebol-Tomas}, P. and {Navarro Umpierrez}, S. and {Pelaez-Torres}, A. and {Reyes Rodriguez}, E. and {Sosa-Guillen}, P. and {Vielba Trillo}, A. and {Esparza-Borges}, E.},
        title = "{Recurrent Nova M31N 2008-12a: discovery of the 2022 eruption with the Las Cumbres Observatory global telescope network}",
      journal = {The Astronomer's Telegram},
         year = 2022,
        month = dec,
       volume = {15786},
        pages = {1},
       adsurl = {https://ui.adsabs.harvard.edu/abs/2022ATel15786....1P}
}

@ARTICLE{Sha22,
       author = {{Shafter}, A.~W. and {Burris}, W.~A. and {Horst}, J.~C. and {Quimby}, R.~M. and {Bardwell}, J. and {Markham-Adkison}, JT and {Thornton}, M. and {Fidrich}, R. and {Darnley}, M.~J.},
        title = "{Multicolor Photometry Near Maximum Light of the 2022 Eruption of the Recurrent Nova M31N 2008-12a}",
      journal = {The Astronomer's Telegram},
         year = 2022,
        month = dec,
       volume = {15797},
        pages = {1},
       adsurl = {https://ui.adsabs.harvard.edu/abs/2022ATel15797....1S}
}

@ARTICLE{hen18a,
       author = {{Henze}, M. and {Darnley}, M.~J. and {Shafter}, A.~W. and {Kafka}, S. and {Kato}, M. and {Williams}, S.~C. and {et al.}},
        title = "{Recurrent nova M31N 2008-12a: Swift/XRT detection of the 2017 eruption}",
      journal = {The Astronomer's Telegram},
         year = 2018,
        month = jan,
       volume = {11130},
        pages = {1},
       adsurl = {https://ui.adsabs.harvard.edu/abs/2018ATel11130....1H}
}

@ARTICLE{hen18b,
       author = {{Henze}, M. and {Darnley}, M.~J. and {Shafter}, A.~W. and {Kafka}, S. and {Kato}, M. and {Williams}, S.~C. and {et al.}},
        title = "{Recurrent nova M31N 2008-12a: Swift UVOT detection of the 2017 eruption}",
      journal = {The Astronomer's Telegram},
         year = 2018,
        month = jan,
       volume = {11121},
        pages = {1},
       adsurl = {https://ui.adsabs.harvard.edu/abs/2018ATel11121....1H}
}

@ARTICLE{hen18c,
       author = {{Henze}, M. and {Darnley}, M.~J. and {Shafter}, A.~W. and {Kafka}, S. and {Kato}, M. and {et al.}},
        title = "{Recurrent nova M31N 2008-12a: Swift UVOT detection of the 2018 eruption}",
      journal = {The Astronomer's Telegram},
         year = 2018,
        month = nov,
       volume = {12182},
        pages = {1},
       adsurl = {https://ui.adsabs.harvard.edu/abs/2018ATel12182....1H}
}

@ARTICLE{hen18d,
       author = {{Henze}, M. and {Darnley}, M.~J. and {Shafter}, A.~W. and {Kafka}, S. and {Kato}, M. and {Williams}, S.~C. and {et al.}},
        title = "{Recurrent nova M31N 2008-12a: Swift/XRT detection of the 2018 eruption}",
      journal = {The Astronomer's Telegram},
         year = 2018,
        month = nov,
       volume = {12207},
        pages = {1},
       adsurl = {https://ui.adsabs.harvard.edu/abs/2018ATel12207....1H}
}

@ARTICLE{dar19a,
       author = {{Darnley}, M.~J. and {Henze}, M. and {Hachisu}, I. and {Kato}, M. and {Osborne}, J.~P. and {Ness}, J. -U. and {Shafter}, A.~W.},
        title = "{Recurrent nova M31N 2008-12a: Swift UVOT detection of the 2019 eruption}",
      journal = {The Astronomer's Telegram},
         year = 2019,
        month = nov,
       volume = {13274},
        pages = {1},
       adsurl = {https://ui.adsabs.harvard.edu/abs/2019ATel13274....1D}
}

@ARTICLE{dar19b,
       author = {{Darnley}, M.~J. and {Henze}, M. and {Hachisu}, I. and {Kato}, M. and {Osborne}, J.~P. and {Ness}, J. -U. and {Shafter}, A.~W.},
        title = "{Recurrent nova M31N 2008-12a: Swift/XRT detection of the 2019 eruption}",
      journal = {The Astronomer's Telegram},
         year = 2019,
        month = nov,
       volume = {13290},
        pages = {1},
       adsurl = {https://ui.adsabs.harvard.edu/abs/2019ATel13290....1D}
}

@ARTICLE{dar20a,
       author = {{Darnley}, M.~J. and {Page}, K.~L.},
        title = "{Recurrent Nova M31N 2008-12a: Swift UVOT detection of the 2020 eruption}",
      journal = {The Astronomer's Telegram},
         year = 2020,
        month = nov,
       volume = {14142},
        pages = {1},
       adsurl = {https://ui.adsabs.harvard.edu/abs/2020ATel14142....1D}
}

@ARTICLE{dar20b,
       author = {{Darnley}, M.~J. and {Page}, K.~L. and {Henz}, M.},
        title = "{Recurrent nova M31N 2008-12a: Swift/XRT detection of the 2020 eruption}",
      journal = {The Astronomer's Telegram},
         year = 2020,
        month = nov,
       volume = {14152},
        pages = {1},
       adsurl = {https://ui.adsabs.harvard.edu/abs/2020ATel14152....1D}
}

@ARTICLE{dar21,
       author = {{Darnley}, M.~J. and {Pag}, K.~L.},
        title = "{Recurrent Nova M31N 2008-12a: Swift UVOT detection of the 2021 eruption}",
      journal = {The Astronomer's Telegram},
         year = 2021,
        month = nov,
       volume = {15040},
        pages = {1},
       adsurl = {https://ui.adsabs.harvard.edu/abs/2021ATel15040....1D}
}

@ARTICLE{dar21a,
       author = {{Darnley}, M.~J. and {Pag}, K.~L.},
        title = "{Recurrent nova M31N 2008-12a: Swift/XRT detection of the 2021 eruption}",
      journal = {The Astronomer's Telegram},
         year = 2021,
        month = nov,
       volume = {15050},
        pages = {1},
       adsurl = {https://ui.adsabs.harvard.edu/abs/2021ATel15050....1D}
}

@ARTICLE{dar22,
       author = {{Darnley}, M.~J. and {Page}, K.~L. and {Healy}, M.~W.},
        title = "{Recurrent nova M31N 2008-12a: Swift/XRT and Swift/UVOT detections of the 2022 eruption}",
      journal = {The Astronomer's Telegram},
         year = 2022,
        month = dec,
       volume = {15798},
        pages = {1},
       adsurl = {https://ui.adsabs.harvard.edu/abs/2022ATel15798....1D}
}

@ARTICLE{whi95,
       author = {{White}, N.~E. and {Giommi}, P. and {Heise}, J. and {Angelini}, L. and {Fantasia}, S.},
        title = "{RX J0045.4+4154: A Recurrent Supersoft X-Ray Transient in M31}",
      journal = {\apjl},
         year = 1995,
        month = jun,
       volume = {445},
        pages = {L125},
          doi = {10.1086/187905},
       adsurl = {https://ui.adsabs.harvard.edu/abs/1995ApJ...445L.125W}
}

@ARTICLE{wil04,
       author = {{Williams}, Benjamin F. and {Garcia}, Michael R. and {Kong}, Albert K.~H. and {Primini}, Frank A. and {King}, A.~R. and {Di Stefano}, Rosanne and {Murray}, Stephen S.},
        title = "{A Synoptic X-Ray Study of M31 with the Chandra High Resolution Camera}",
      journal = {\apj},
         year = 2004,
        month = jul,
       volume = {609},
       number = {2},
        pages = {735-754},
          doi = {10.1086/421315},
archivePrefix = {arXiv},
       eprint = {astro-ph/0306421},
 primaryClass = {astro-ph},
       adsurl = {https://ui.adsabs.harvard.edu/abs/2004ApJ...609..735W}
}

@ARTICLE{bar11,
       author = {{Barsukova}, E. and {Fabrika}, S. and {Hornoch}, K. and {Fatkhullin}, T. and {Sholukhova}, O. and {Pietsch}, W.},
        title = "{Spectroscopy of nova 2011-10d and photometry of nova candidate 2011-10e in M31}",
      journal = {The Astronomer's Telegram},
         year = 2011,
        month = oct,
       volume = {3725},
        pages = {1},
       adsurl = {https://ui.adsabs.harvard.edu/abs/2011ATel.3725....1B}
}

@ARTICLE{sha12,
       author = {{Shafter}, A.~W. and {Hornoch}, K. and {Ciardullo}, J. Vrastil R. and {Darnley}, M.~J. and {Bode}, M.~F.},
        title = "{Spectroscopic Observations of the Unusual Transient TCP J00452884+4154095 in M31}",
      journal = {The Astronomer's Telegram},
         year = 2012,
        month = oct,
       volume = {4503},
        pages = {1},
       adsurl = {https://ui.adsabs.harvard.edu/abs/2012ATel.4503....1S}
}

@ARTICLE{tan13,
       author = {{Tang}, S. and {Cao}, Y. and {Kasliwal}, M.~M.},
        title = "{A Candidate Recurrent Nova in M31 from iPTF}",
      journal = {The Astronomer's Telegram},
         year = 2013,
        month = nov,
       volume = {5607},
        pages = {1},
       adsurl = {https://ui.adsabs.harvard.edu/abs/2013ATel.5607....1T}
}

@ARTICLE{dar15a,
       author = {{Darnley}, M.~J. and {Henze}, M. and {Shafter}, A.~W. and {Kato}, M.},
        title = "{Recurrent Nova M31N 2008-12a: discovery of the 2015 eruption}",
      journal = {The Astronomer's Telegram},
         year = 2015,
        month = aug,
       volume = {7964},
        pages = {1},
       adsurl = {https://ui.adsabs.harvard.edu/abs/2015ATel.7964....1D}
}

@ARTICLE{dar15b,
       author = {{Darnley}, M.~J. and {Henze}, M. and {Shafter}, A.~W. and {Kato}, M.},
        title = "{Recurrent Nova M31N 2008-12a: spectroscopic confirmation of the 2015 eruption and additional photometry}",
      journal = {The Astronomer's Telegram},
         year = 2015,
        month = aug,
       volume = {7965},
        pages = {1},
       adsurl = {https://ui.adsabs.harvard.edu/abs/2015ATel.7965....1D}
}

@ARTICLE{hen15b,
       author = {{Henze}, M. and {Darnley}, M.~J. and {Shafter}, A.~W. and {Kato}, M. and {Hachisu}, I. and {Bode}, M.~F. and {Ness}, J. -U. and {Osborne}, J.~P. and {Kennea}, J.~A. and {Gehrels}, N.},
        title = "{Recurrent nova M31N 2008-12a: the 2015 outburst detected in X-rays with Swift XRT}",
      journal = {The Astronomer's Telegram},
         year = 2015,
        month = sep,
       volume = {7984},
        pages = {1},
       adsurl = {https://ui.adsabs.harvard.edu/abs/2015ATel.7984....1H}
}

@ARTICLE{ita16a,
       author = {{Itagaki}, K. and {Gao}, X. and {Darnley}, M.~J. and {Henze}, M. and {Shafter}, A.~W. and {Williams}, S.~C. and {Kafka}, S. and {Kato}, M. and {et al.}},
        title = "{Recurrent Nova M31N 2008-12a: discovery and constraints of the 2016 eruption}",
      journal = {The Astronomer's Telegram},
         year = 2016,
        month = dec,
       volume = {9848},
        pages = {1},
       adsurl = {https://ui.adsabs.harvard.edu/abs/2016ATel.9848....1I}
}

@ARTICLE{hen16,
       author = {{Henze}, M. and {Darnley}, M.~J. and {Shafter}, A.~W. and {Kafka}, S. and {Kato}, M. and {et al.}},
        title = "{Recurrent Nova M31N 2008-12a: Swift UVOT detection of the 2016 eruption}",
      journal = {The Astronomer's Telegram},
         year = 2016,
        month = dec,
       volume = {9853},
        pages = {1},
       adsurl = {https://ui.adsabs.harvard.edu/abs/2016ATel.9853....1H}
}

@ARTICLE{hen16a,
       author = {{Henze}, M. and {Darnley}, M.~J. and {Shafter}, A.~W. and {Kafka}, S. and {Kato}, M. and {et al.}},
        title = "{Recurrent nova M31N 2008-12a: Swift/XRT detection of the 2016 eruption}",
      journal = {The Astronomer's Telegram},
         year = 2016,
        month = dec,
       volume = {9872},
        pages = {1},
       adsurl = {https://ui.adsabs.harvard.edu/abs/2016ATel.9872....1H}
}

@ARTICLE{gal20,
       author = {{Galloway}, D.~K. and {Ackley}, K. and {Wiersema}, K. and {Dyer}, M. and {Lyman}, J. and {Ulaczyk}, K. and {Steeghs}, D. and {Dhillon}, V. and {O'Brien}, P. and {Ramsay}, G. and {Poshyachinda}, S. and {Kotak}, R. and {Nuttall}, L. and {Pollacco}, D. and {Breton}, R. and {GOTO Collaboration}},
        title = "{GOTO detection of the M31 nova M31N 2008-12a}",
      journal = {The Astronomer's Telegram},
         year = 2020,
        month = nov,
       volume = {14141},
        pages = {1},
       adsurl = {https://ui.adsabs.harvard.edu/abs/2020ATel14141....1G}
}

@MISC{heasoft14,
       author = {{Nasa High Energy Astrophysics Science Archive Research Center (Heasarc)}},
        title = "{HEAsoft: Unified Release of FTOOLS and XANADU}",
 howpublished = {Astrophysics Source Code Library, record ascl:1408.004},
         year = 2014,
        month = aug,
          eid = {ascl:1408.004},
        pages = {ascl:1408.004},
archivePrefix = {ascl},
       eprint = {1408.004},
       adsurl = {https://ui.adsabs.harvard.edu/abs/2014ascl.soft08004N}
}

@inproceedings{mckinney10,
  title={Data structures for statistical computing in python},
  author={McKinney, Wes and others},
  booktitle={Proceedings of the 9th Python in Science Conference},
  volume={445},
  pages={51--56},
  year={2010},
  organization={Austin, TX}
}

@Article{Hunter07,
  Author    = {Hunter, J. D.},
  Title     = {Matplotlib: A 2D graphics environment},
  Journal   = {Computing in Science \& Engineering},
  Volume    = {9},
  Number    = {3},
  Pages     = {90--95},
  publisher = {IEEE COMPUTER SOC},
  doi       = {10.1109/MCSE.2007.55},
  year      = 2007
}

@ARTICLE{Kato20,
       author = {{Kato}, Mariko and {Hachisu}, Izumi},
        title = "{Supersoft X-ray phases of recurrent novae as an indicator of their white dwarf masses}",
      journal = {\pasj},
         year = 2020,
        month = oct,
       volume = {72},
       number = {5},
          eid = {82},
        pages = {82},
          doi = {10.1093/pasj/psaa071},
archivePrefix = {arXiv},
       eprint = {2007.03387},
 primaryClass = {astro-ph.SR},
       adsurl = {https://ui.adsabs.harvard.edu/abs/2020PASJ...72...82K}
}
\bibliographystyle{aasjournal}

\end{document}